\documentclass[prd,twocolumn,showpacs,amsmath,amssymb,citeautoscript]{revtex4}

\usepackage{graphicx,times}
\usepackage{amsmath,amssymb,amsbsy,amsfonts}
\usepackage{array}
\usepackage{bm}
\usepackage{graphics}
\usepackage{mathrsfs}
\usepackage{subfig}

\newcommand{\psib}{{\overline{\psi}}}

\begin{document}

\title{Generating a non-perturbative mass gap using Feynman diagrams in an asymptotically free theory}
\author{Venkitesh Ayyar}
\affiliation{Department of Physics, University of Colorado, Boulder, Colorado 80309, USA}
\author{Shailesh Chandrasekharan}
\affiliation{Department of Physics, Box 90305, Duke University,
Durham, North Carolina 27708, USA}

\begin{abstract}
Using the example of a two dimensional four-fermion lattice field theory we demonstrate that Feynman diagrams can generate a mass gap when massless fermions interact via a marginally relevant coupling. We introduce an infrared cutoff through the finite system size so that the perturbation series for the partition function and observables become convergent. We then use the Monte Carlo approach to sample sufficiently high orders of diagrams to expose the presence of a mass gap in the lattice model.
\end{abstract}


\maketitle

\section{Introduction}

Understanding how a mass gap is generated in an asymptotically free theory like Yang-Mills theory continues to be a fascinating topic of research. Using Wilson's lattice formulation the origin of the mass gap is easy to derive within the strong coupling expansion \cite{Kogut:1979wt}. Monte Carlo calculations have shown that the mass gap continues to exist and scales appropriately even for much weaker couplings. However, the challenge of course is to begin with a weak coupling expansion and show the presence of the mass gap. A Monte Carlo method that directly works within the weak coupling expansion could perhaps shed more light on the subject.

Recently, Monte Carlo methods have emerged that sample weak coupling Feynman diagrams in a variety of models  \cite{Pro98,PhysRevLett.99.250201,Bon06,PhysRevB.77.125101,Kozik:2010zz,PhysRevLett.115.266802}. Can such methods also be applicable to asymptotically free theories like Yang Mills theories and QCD?  The obvious problem is that the weak coupling approach is an expansion in powers of the coupling $g$, while mass gaps in these theories arise non-perturbatively through an essential singularity of the form $M \sim \mathrm{e}^{-\beta/g^2}$. So, at least naively, it seems impossible that weak coupling diagrams can be combined with Monte Carlo methods to generate a mass gap. As a first step in addressing this impasse, one can even ignore complications of a gauge theory and ask whether these weak coupling approaches can generate a mass gap in simpler two dimensional spin models that are known to be asymptotically free. This question was raised recently and partially addressed within the context of the two dimensional $O(N)$ and $U(N)\times U(N)$ model in the large $N$ limit \cite{Buividovich:2015qba,Buividovich:2017sdn}. The strategy that seems to work is to regulate the infrared divergences in a controllable way so as to make the weak coupling series convergent. A resummation of the convergent series then exposes the existence of the mass gap. 

In this work we use a similar strategy but consider a $SU(4)$ symmetric four-fermion model and thus explore the question of whether Feynam diagrams can generate a non-perturbative mass gap in asymptotically free theories without the simplifications that usually occur at large $N$ \cite{Kopper:1993mj,Kopper:1999jj}. Two dimensional four-fermion field theories can be asymptotically free  \cite{PhysRevD.10.3235,Witten1978110}, and can be formulated to have completely convergent weak coupling expansion by formulating them on a finite space-time lattice.
However, in the absence of a small parameter like $1/N$ in large $N$ models, the weak coupling diagrammatic series may converge only after summing over many terms. In our work we accomplish this summation using a Monte Carlo method and hence are able still expose the presence of a non-perturbative mass gap that is independent of the infrared regulator. By tuning the bare coupling to zero we can also explore the continuum limit. From a continuum quantum field theory perspective, there are connections of our approach to recent ideas of using resurgent functions and trans-series combined with boundary conditions that control infrared divergences to define the perturbation series non-perturbatively \cite{Dunne:2012ae,Dunne:2013ada,Cherman:2013yfa}.

The physics of our lattice model is interesting from other perspectives as well. For example it was recently studied extensively in three and four dimensions \cite{PhysRevD.93.081701,PhysRevD.91.065035,Catterall2016,Ayyar2016,PhysRevD.96.034506} and contains a weak coupling massless fermion phase and a strong coupling massive fermion phase without any spontaneous symmetry breaking.  In three dimensions one finds a second order phase transition that separates these two phases. This quantum critical point is exotic and may contain emergent gauge fields \cite{You:2017ltx}. We believe this critical point moves to the origin in two dimensions, and thus the mass generation mechanism in our model is likely to be similar to the one discussed in \cite{BenTov:2014eea}. However, the mass should still scale exponentially with the coupling as in any asymptotically free theory. We show this explicitly in our work.

\section{Lattice Model}

Two dimensional lattice four-fermion models have been studied extensively using Monte Carlo methods in the past, but mostly within the Wilson fermion formulation \cite{NAGAI2006325,PhysRevD.76.014503,Bietenholz:1994ur,Ichinose:1999rr,Aoki:1985jj}. The most efficient way to perform the calculations involve using the worldline representation \cite{Bar:2009yq,Korzec:2006hy}. However, this representation is not helpful for understanding how weak coupling perturbation theory using Feynman diagrams can help to generate the non-perturbative mass gap in these models. A simple model that can be studied by sampling Feynman diagrams using a Monte Carlo method, is the reduced staggered fermion model whose action is given by 
\begin{equation}
S(\psi) \ =\ \frac{1}{2}\sum_{x,y,a}\ \psi^a_x \ M_{x,y} \ \psi^a_y \  - \ U\ \sum_x \ \psi^4_x \psi^3_x \psi^2_x \psi^1_x.
\label{act}
\end{equation}
where $M_{x,y}$ is the free staggered fermion matrix 
\begin{equation}
M_{x,y} \ =\ \frac{1}{2}\ \sum_{\alpha} \ \eta_{\alpha,x}\ \big(\delta_{x+\hat{\alpha},y} -  \delta_{x-\hat{\alpha},y}\big).
\end{equation}
with the phase factors $\eta_{1,x}=1$, $\eta_{2,x}=(-1)^{x_1}$. It is easy to verify that this model can be obtained by discretizing the continuum two dimensional model,
\begin{eqnarray}
&& S_{\rm cont} = \int d^2x \ \Big\{\ \sum_{a=1,2; i=1,2} 
\psib^i_a(x)(\sigma_\alpha)_{ij}\partial_\alpha \psi^j_a(x)  \nonumber \\ 
&& \hskip-0.2in
-\  U\Big(\psi^2_1(x) \psi^1_1(x) \psi^2_2(x) \psi^1_2(x) + \psib^2_1(x) \psib^1_1(x) \psib^2_2(x) \psib^1_2(x) 
\Big) \Big\},\nonumber \\
\label{contact}
\end{eqnarray}
where $\sigma_\alpha$ are $2 \times 2$ Pauli matrices. Discretizing (\ref{contact}) naively on a space-time lattice and using the well known spin diagonalization transformation in order to reduce the fermion doubling \cite{Sharatchandra:1981si,vandenDoel:1983mf}, we obtain (\ref{act}).

Note that there are no $\psib^a_x$ fields in the lattice action. In the reduced staggered formulation, one keeps only the minimimal number of fermion fields per site and defines them as $\psi^a_x$ on all sites. We can define the partition function of our model to be
\begin{equation}
Z = Z_0 \int [d\psi] \ \mathrm{e}^{-S(\psi)}
\end{equation}
where $Z_0$ is chosen so that $Z=1$ in the free theory. The Grassmann integration measure $[d\psi]$ is a product of $d\psi^1_x d\psi^2_x d\psi^3_x d\psi^4_x$ on every site $x$.

\begin{figure}[h]
\begin{center}
\includegraphics[width=0.45\textwidth]{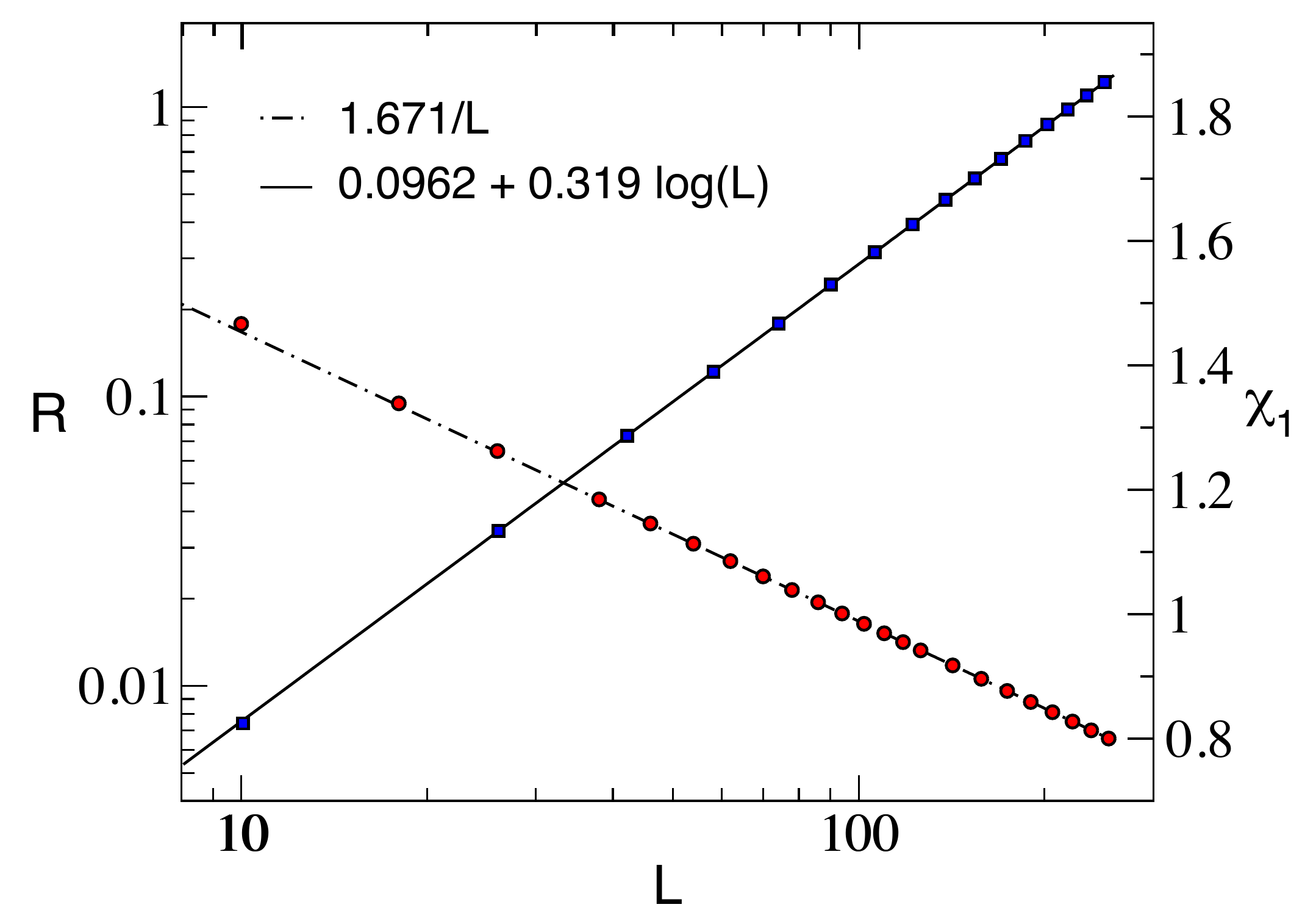}
\end{center}
\caption{\label{fig1} The scaling of the fermion propagator at the mid point ($R$) and the susceptibility ($\chi_1$) as a function of $L$ in the free theory. This shows that the coupling $U$ is perturbatively marginal as expected in the continuum.}
\end{figure}

At $U=0$ our lattice model describes $N_f=4$ flavors of free massless (two-component) Dirac fermions in the continuum limit, which is reached by simply exploring physics at large length scales as compared to the lattice spacing. As a probe of the long distance physics we can take space-time to be a torus of side $L$ (in lattice units) in each direction with anti-periodic boundary conditions. In two dimensions, a free fermion field is expected to have a mass dimension $[\psi^i_a] = 1/2$. In our approach this can be seen by the scaling of the fermion propagator $G_f(x,y)\sim 1/|x-y|$ for large separations. In Fig.~\ref{fig1} we plot the scaling of the propagator at a separation of $|x-y| = L/2$ along one of the directions, $R = G_f(0,L/2)$ as a function of $L$ and find that $R \sim 1.67/L$. In the same figure we also show the scaling of the susceptibility
\begin{equation}
\chi_1 = \frac{1}{2 Z} \int [d\psi] \mathrm{e}^{-S} \ 
\sum_y \Big\{\psi^a_0\psi^b_0 \psi^b_y \psi^a_y\Big\},
\label{sus}
\end{equation}
as a function of $L$ and see the expected infrared divergence.  This means the coupling $U$ is marginal as expected from continuum perturbative power counting. We will see later that in fact $U$ is marginally relevant since an exponentially small mass gap is generated in this model when $U > 0$. This is consistent with 
asymptotic freedom as predicted originally by Gross and Neveu \cite{PhysRevD.10.3235}.

\section{The Partition Function}

The partition function of our model can be expanded in powers of the coupling
\begin{equation}
Z \ =\ \sum_k \ z_k U^k\ =\ \sum_k \Big(\sum_{[x;k]} \Omega([x;k])\Big)\ U^k
\label{zexp}
\end{equation}
where the coefficients $z_k$ can be obtained as a sum over vertex configurations $[x;k] = \{x_1,x_2,...,x_k\}$ of an ordered set of $k$ different lattice sites where the interactions occur. The weight of each vertex configuration is given by 
\begin{equation}
\Omega([x;k]) \ =\ Z_0\ \Big(\int [\prod_xd\psi_x] \mathrm{e}^{-\frac{1}{2} \psi_x \ M_{x,y} \ \psi_y} \ \psi_{x_1} \psi_{x_2}...\psi_{x_k} \Big)^4,
\label{bwt}
\end{equation}
which is a sum over Feynman diagrams obtained through the regular Wick contractions. For each flavor the sum gives the Pfaffian of a $k \times k$ matrix $W([x;k])$, whose matrix elements are given by the free staggered fermion propagator $G_f(x_i,x_j)$ between the sites in $[x;k]$ \cite{Ayyar2016}. Thus, we obtain
\begin{equation}
\Omega([x;k]) \ =\ \Big(\mathrm{Pf}(W([x;k])\Big)^4,
\label{om}
\end{equation}
which is guaranteed to be positive. A Monte Carlo method can then be used to sample vertex configurations $[x;k]$ distributed according to the probability distribition 
\begin{equation}
P_k(U,[x;k]) \ =\ \frac{U^k}{Z(U)} \Big(\mathrm{Pf}(W([x;k])\Big)^4.
\label{om}
\end{equation}
Due to symmetries of the model the only non-zero weights involve an equal number of even and odd sites in $[x;k]$ which implies that only even values of $k$ contribute to the expansion. 

\begin{figure}[h]
\begin{center}
\includegraphics[width=0.45\textwidth]{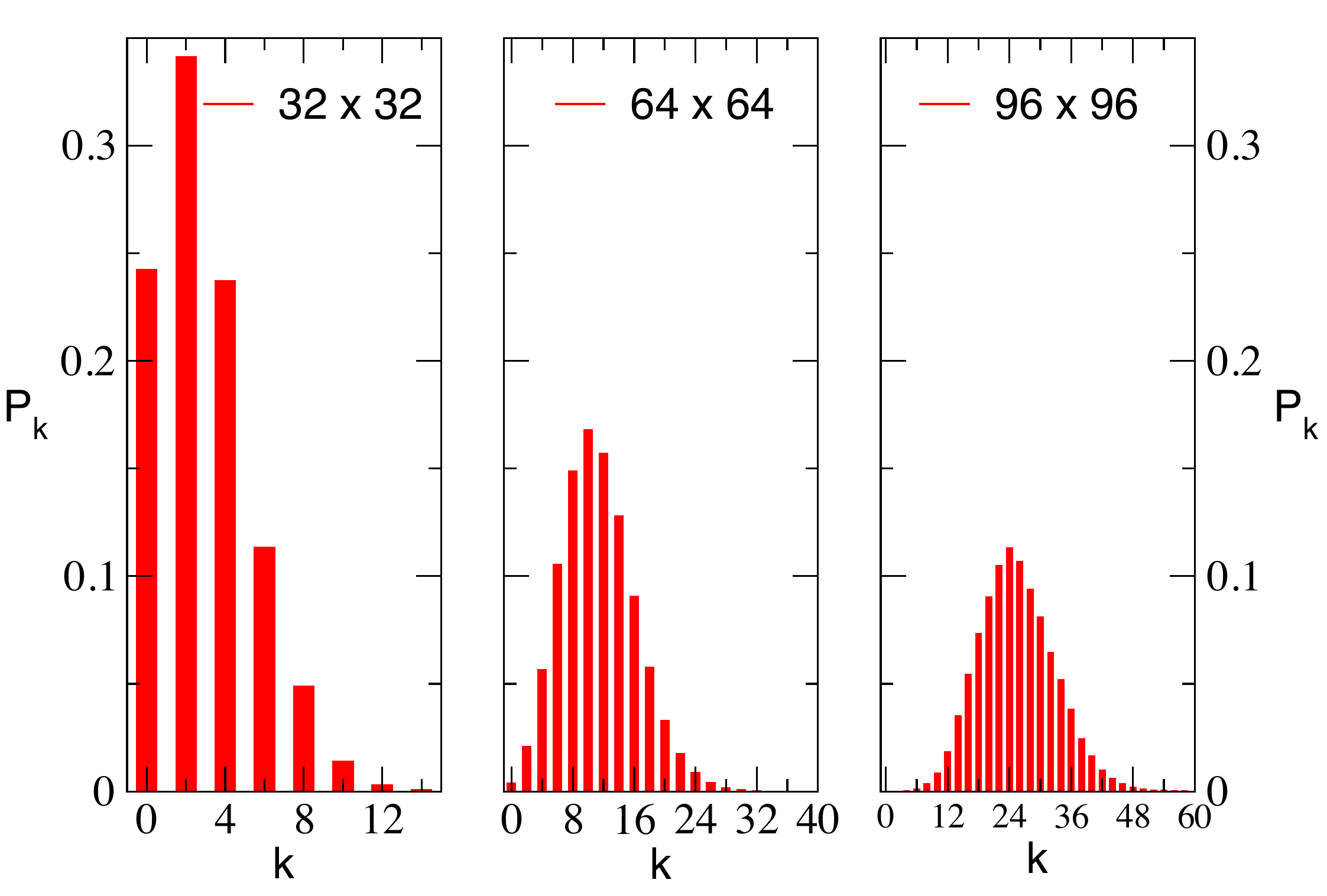}
\end{center}
\caption{\label{fig2} Probability distribution of vertices in the partition function as a function of the lattice size at $U=0.1$ for $L=32,64$ and $96$. The average density of vertices $\rho_k = \langle k\rangle/L^2 \approx 0.00267$ remains constant in all the three cases.}
\end{figure}

\begin{figure}[h]
\begin{center}
\includegraphics[width=0.45\textwidth]{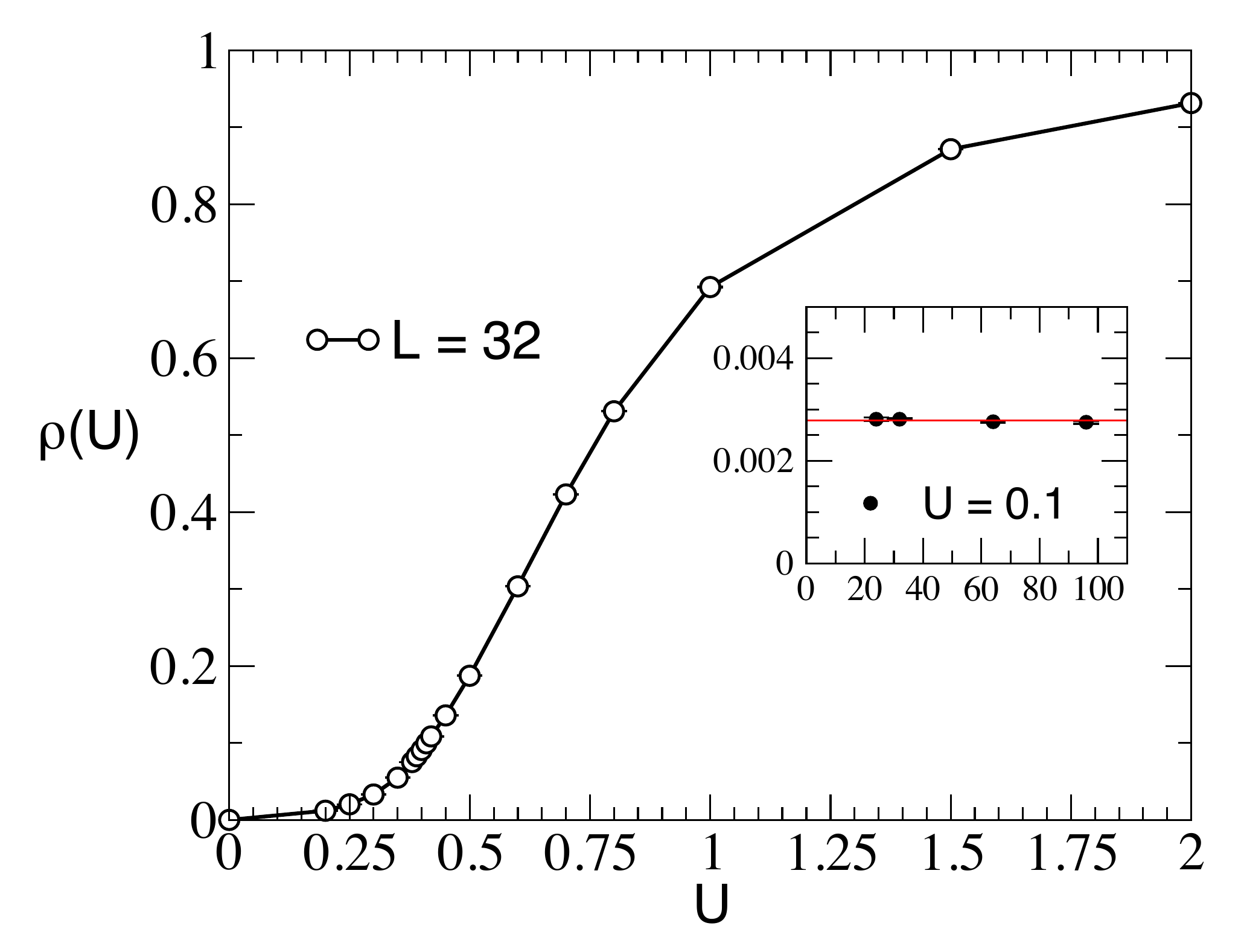}
\end{center}
\caption{\label{fig4} Plot of the density of vertices $\rho(U)$ as a function of $U$. The inset shows the density at $U=0.1$ as a function of $L$. We see that the density of vertices remains the same as $L$ increases.}
\end{figure}

Interestingly, the expansion (\ref{zexp}) in powers of $U$ is finite and completely convergent on a finite lattice, since the maximum number of vertices that are allowed is $L^2$. Our goal is to understand how the infrared divergences present in an asymptotically free theory manifest themselves in this convergent expansion. In order to gain some insight into the dominant terms in the expansion we define the probability distribution of vertices, $P_k(U)=z_kU^k/Z(U)$. Note that $P_k(U)$ as the sum over $P_k(U,[x,k])$ with a fixed $k$ and $U$ but different locations of the vertices. We can use Monte Carlo sampling to compute $P_k(U)$. In Fig.~\ref{fig2} we plot this probability distribution of vertices at $U=0.1$ for different values of $L$. As we can see, sectors with large number of vertices are suppressed exponentially and the average number of vertices is much smaller than the maximum value $k_{\rm max}=L^2$. We discover that a more useful quantity is the average density of vertices $\rho(U) = \langle k\rangle/L^2$. In Fig.~\ref{fig4} we show how $\rho(U)$ changes with $U$. In the inset we plot this density at $U=0.1$ for various lattice sizes and observe that it rapidly saturates in the thermodynamic limit. At $U=0.1$, the average density is very small $\rho = 0.0027$.

\begin{figure}[htb]
\begin{center}
\includegraphics[width=.45\textwidth]{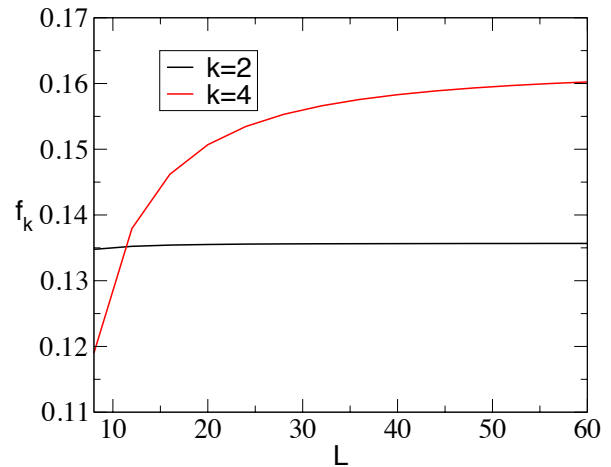}
\end{center}
\caption{\label{fig3} Plot of the perturbative coefficients $f_2$ and $f_4$ in the expansion of the free energy as a function of $L$.}
\end{figure}

It is easy to understand why the average density of vertices approaches a constant in the thermodynamic limit. From a statistical mechanics point of view one expects that the partition function scales as $Z = \exp( f(U) L^2)$ in the thermodynamic limit, where $f(U)$ is the free energy density. Since $f(U)$ is expected to be independent of the volume for sufficiently large volumes, the average density of vertices
\begin{equation}
\rho(U) = \frac{\langle k\rangle}{L^2} \ =\ (U/L^2) (\partial \ln Z(U)/\partial U)\ =\ 
U (\partial f(U)/\partial U),
\end{equation} 
is also independent of $L$. At a fixed $L$ we can expand $f(U) = f_2 U^2 + f_4 U^4+...$ and find connections between $f_k$'s and $z_k$'s. For example $f_2 = z_2/L^2$ and $f_4 = (z_4 - z_2^2/2)/L^2$ and so on. In Fig.~\ref{fig3} we plot $f_2$ and $f_4$ as functions of $L$ for our model and see that both these coefficients are well behaved and do not show infrared divergences. 

The connection between $Z(U)$ and $f(U)$ is well known in diagrammatic perturbation theory; the former contains contributions from disconnected diagrams, while the latter
gets contributions only from connected diagrams. If there are infrared divergences in perturbation theory they could appear in the $f_k$ coefficients. From the discussion above we have shown that $Z(U)$ contains no infrared divergences except for the usual factors of the volume, but we cannot rule out such divergences in the expansion of $f(U)$, although in our model they do not appear in the first two terms $f_2$ and $f_4$. This seems to be a feature of our current model due to its symmetries. Even if $f_k$'s contained divergences we can still extract $f(U)$ non-perturbatively through the integral
\begin{equation}
f(U) \ = \ \int_0^U \rho(U)/U.
\end{equation}
However, we will need to compute $\rho(U)$ non-perturbatively by summing over the distribution of vertices generated by the Monte Carlo method. The usual infrared divergences in perturbation theory disappear once this resummation is performed.

\section{The Mass Gap}

\begin{figure}[htb]
\begin{center}
\includegraphics[width=.45\textwidth]{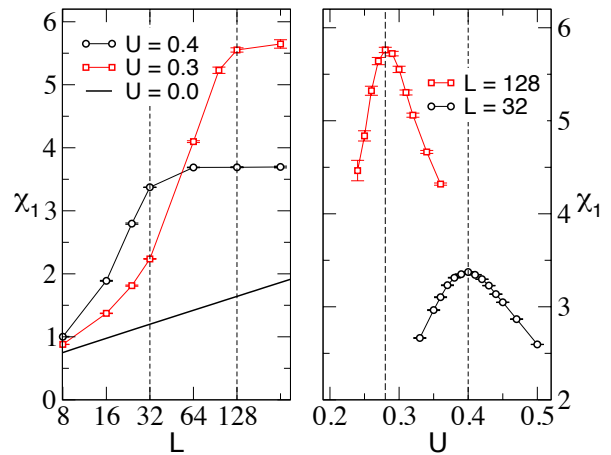}
\end{center}
\caption{\label{fig5} Plot of the susceptibility $\chi_1$ as a function of $U$ for different lattice sizes. For value of $L$, we can define the mass scale $M_b' = 1/L$ that is generated when the location of the peak $U=U_{\rm p}$ determines the scale $M_b' = 1/L$.}
\end{figure}

In order to see how the diagrammatic method reproduces the mass gap we have studied two observables that are sensitive to the mass gap and both give very similar results \cite{Suppmat}. Here we focus on one of them, which is the finite size susceptibility $\chi_1$ defined in (\ref{sus}). As we already pointed out in Fig.~\ref{fig1}, in the free theory $\chi_1$ diverges logarithmically for large values of the lattice size $L$. However, if the mass gap $M$ is generated we expect $\chi_1$ will begin to level off roughly around $L \sim 1/M$. Further, the calculation of $\chi_1$ can also be expressed as a sum over Feynman diagrams through the relation,
\begin{equation}
\chi_1 \ =\ \sum_{y,k} \Big(\sum_{[x;k]} \Gamma_{0,y}([x;k]) \ P_k(U,[x;k]) \Big)
\end{equation}
where $\Gamma_{0,y}([x;k])$ is the ratio of two quantities: the numerator is the sum over all Feynman diagrams with two external sources located at the origin and at $y$ in addition to the configuration of interaction vertices  $[x;k] = \{x_1,x_2,...,x_k\}$ and the denominator is $\Omega([x;k])$, i.e., the sum over Feynman diagrams without the sources. This division makes $\Gamma_{0,y}([x;k])$ scale like a ``connected'' Feynman diagram for large volumes since a factor that scales exponentially in the volume is cancelled in the ratio. The configuration of interaction vertices $[x;k]$ is generated with probability $P_k(U,[x;k])$ and $\chi_1$ is measured by choosing a lattice site at random which is defined as the origin and summing over all possible locations of $y$. If $\chi_1$ contains infrared divergences, then it will increase indefinitely with $L$. We know that at $U=0$ this does indeed occur as shown in Fig.~\ref{fig1}.  On the other hand, in our asymptotically free theory we expect a non-perturbative mass gap $M \sim \exp(-\beta/U)$ to be generated and $\chi_1$ to level off when $L > 1/M$. In the left figure of Fig.~\ref{fig5} we plot $\chi_1$ as a function of $L$ at $U=0.3$ and $0.4$. We observe that indeed $\chi_1$ begins to level off around $L \sim 128$ at $U=0.3$ and around $L \sim 32$ at $U=0.4$. The fact that it takes a substantially larger lattice to level off at $U=0.3$ as compared to $U=0.4$ is an indication that $M$ is decreasing rapidly. We also plot the $U=0$ results for comparison. 

\begin{figure}[htb]
\begin{center}
\includegraphics[width=.4\textwidth]{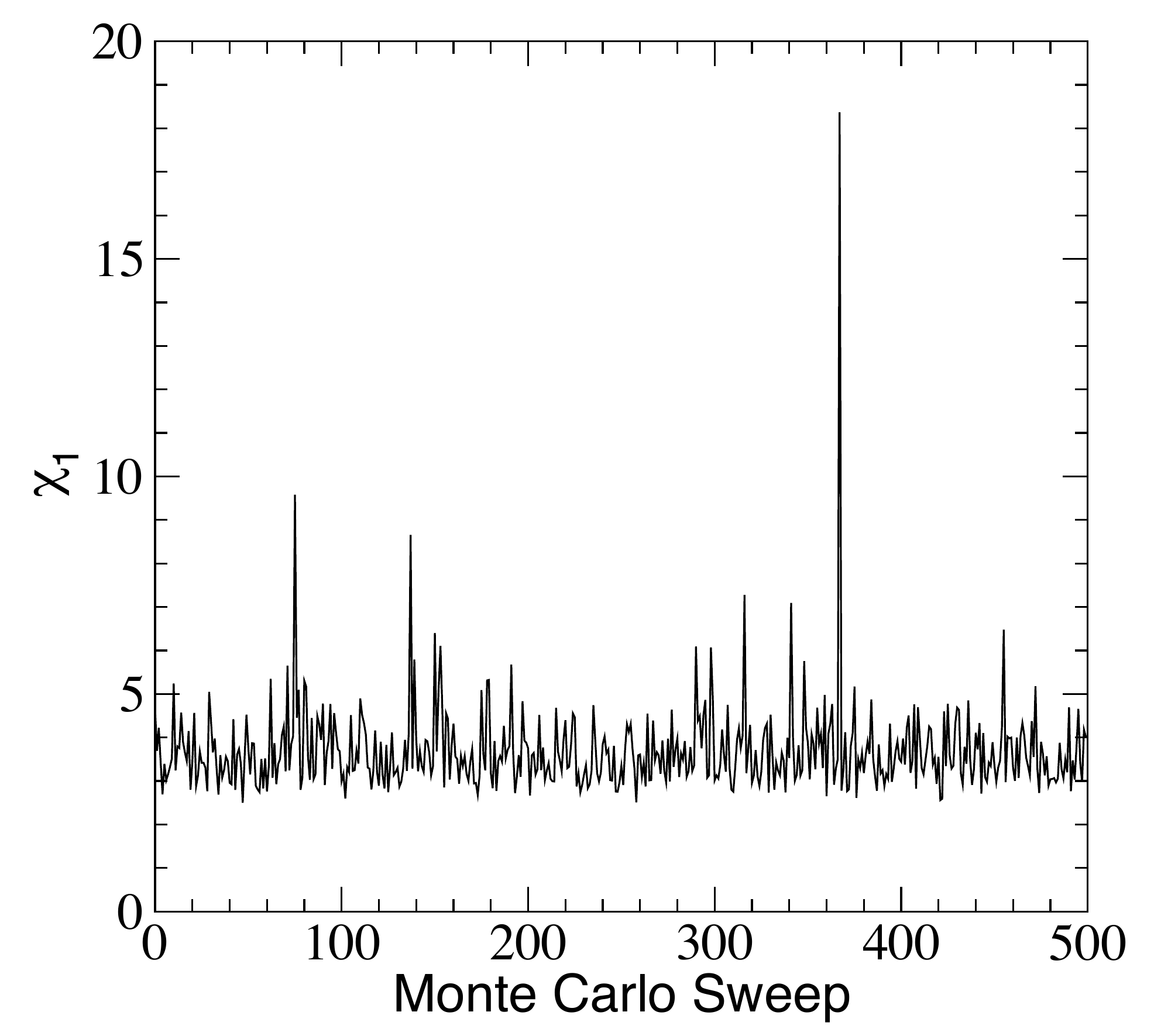}
\end{center}
\caption{\label{fig8} Fluctuations of $\chi_1$ in a sample of $500$ vertex configurations generated consecutively during the Monte Carlo sampling.}
\end{figure}

Statistically speaking this implies that for most vertex configurations $[x;k]$ generated in the Monte Carlo sample, $\Gamma_{0,y}([x;k])$ begins to decay exponentially for points $y$ far from the origin, although when $y$ is close to the origin there is some enhancement when $U > 0$ as compared to $U=0$ (see left figure of Fig.~\ref{fig5}).
It seems that the infrared divergence of the usual perturbation theory disappears for large lattices when we take into account a constant density of vertices. This means we need to consider large orders of perturbation theory. But what about the infrared divergences that clearly exist at small orders of perturbation theory? We believe these are the ones that cause the enhancement in $\chi_1$ at small values of $L$ but eventually, at large values of $L$, become statistically insignificant. In other words they are rare and hidden in the Monte Carlo fluctuations of the vertices that are generated. To see this, in Fig.~\ref{fig8} we plot the fluctuations in $\chi_1$ during a sample of the Monte Carlo time history for $L=64$ and $U=0.4$. As can be seen from Fig.~\ref{fig5}, for these parameters the theory has generated a mass gap with $\chi_1 \approx 3.7(5)$ being the saturated value of the susceptibility. However, as Fig.~\ref{fig8} shows there are still large but rare fluctuations in $\chi_1$ that are three to four times larger than the average value. In perturbation theory the fact that these logarithmically divergent contributions are rare compared to finite contributions cannot be uncovered easily.

\begin{figure*}[htb!]
\begin{center}
\includegraphics[width=.43\textwidth]{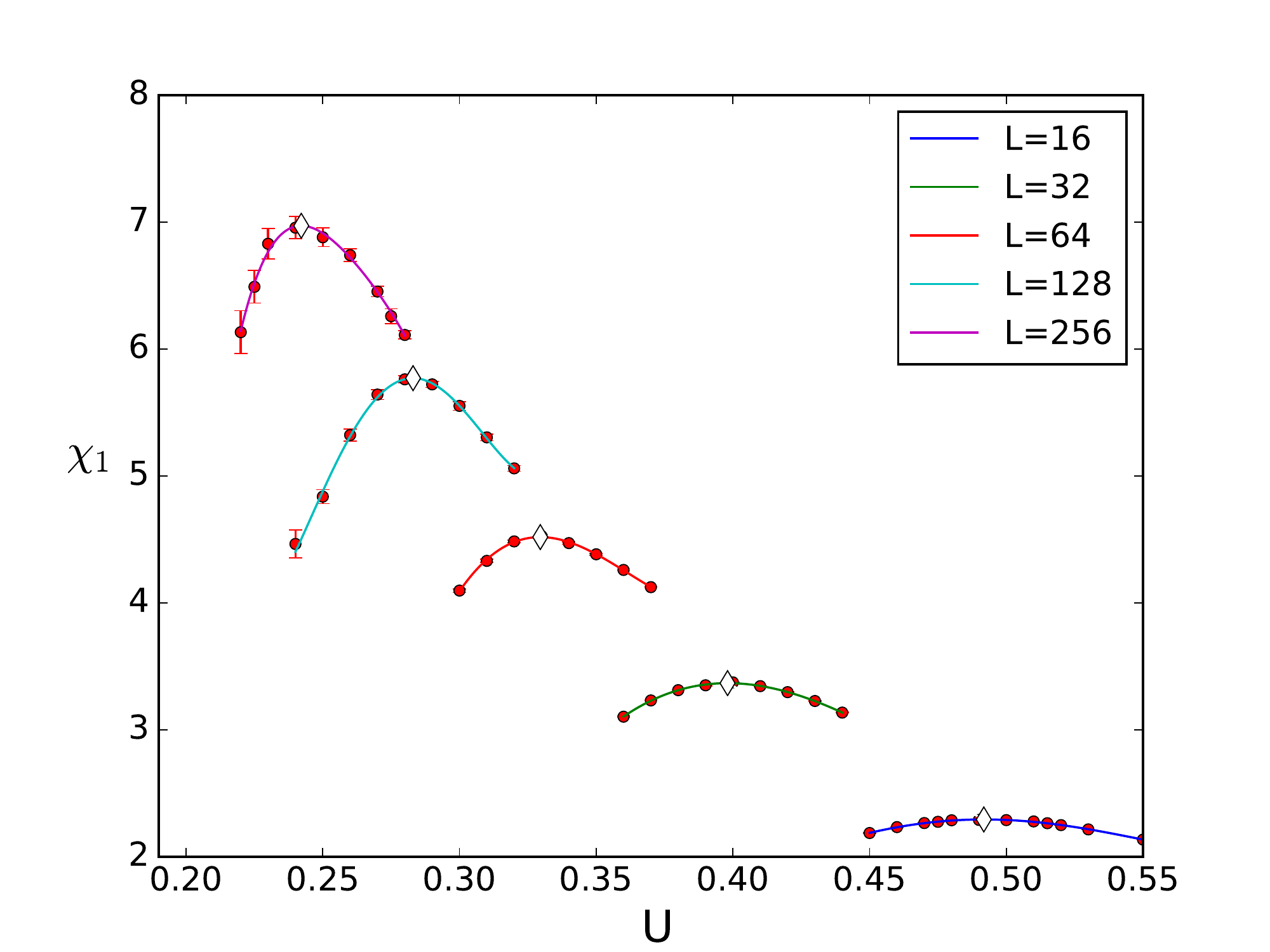}
\hskip0.2in \includegraphics[width=.51\textwidth]{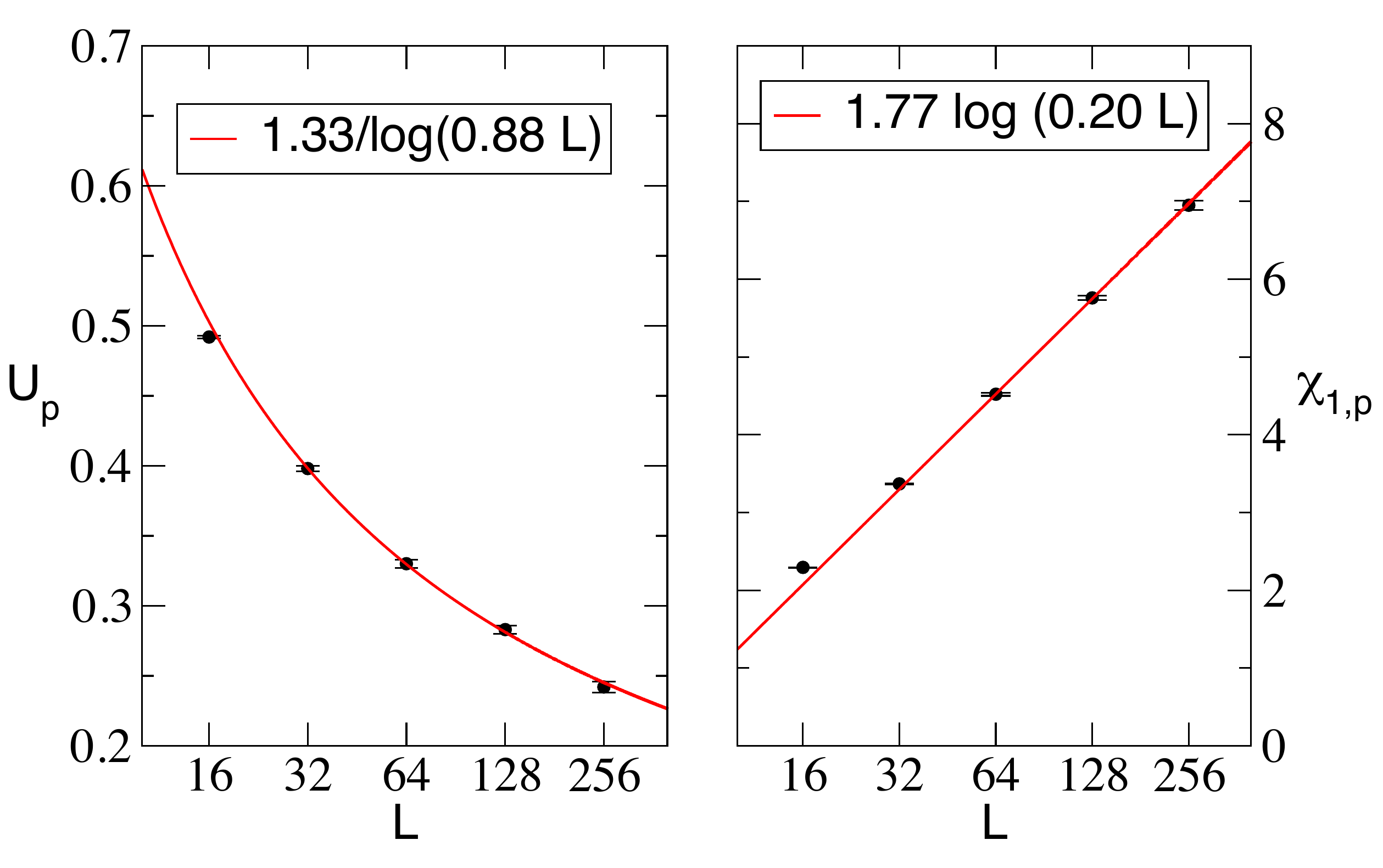}
\end{center}
\caption{\label{fig7} (Left Figure) Plot of $\chi_1$ as a function of $U$ for different values of $L$. The locations of the peak obtained by fitting the data to a smooth curve are listed in Tab.~\ref{tab1}. (Right Figure) Plot of $U_{\rm p}$ and $\chi_{1,{\rm p}}$ as a function of $L$ and the fits to Eq.(\ref{upfit}).}
\end{figure*}

From the left plot of Fig.~\ref{fig5} we roughly expect $M \sim 1/32$ at $U=0.4$ which decreases to $M \sim 1/128$ at $U=0.3$. Using the exponential dependence of $\chi_1(L=\infty) - \chi_1(L)$ on $L$ we can in principle extract $M$ quantitatively. However, here we devise an alternate procedure and determine a slightly different mass scale $M_b$ as follows. We first note that the susceptibility $\chi_1$ has a peak as a function of $U$ for every fixed value of $L$. This behavior is clearly visible in the right figure of Fig.~\ref{fig5} where we plot $\chi_1$ as a function of $U$ at $L=32$ and $128$. For a fixed $L$ if the peak occurs at $U=U_{\rm p}$, we define a non-perturbative mass scale at $U_{\rm p}$ using the relation $M_b \equiv 1/L$. Comparing the left and right figures in Fig.~\ref{fig5} we see that our definition of $M_b$ is also roughly consistent with the value of $1/L$ where $\chi_1$ begins to level off. The locations of the peaks at various lattice sizes can be obtained rather accurately by fitting the data as shown in Fig.~\ref{fig7}. Table \ref{tab1} gives the values of the peak obtained through such a fit at different values of $L$.

\begin{table}[h]
\begin{center}
\begin{tabular}{|c|c|c|}
\hline
$L$ & $\chi_{1,{\rm p}}$ & $U_{\rm p}$ \\
\hline
16 & 2.293(2) & 0.492(1) \\
32 & 3.368(5) & 0.398(2) \\
64 & 4.520(20) & 0.330(3) \\
128 & 5.760(30) & 0.283(3) \\
256 & 6.950(60) & 0.242(4) \\
\hline
\end{tabular}
\end{center}
\caption{\label{tab1} Fit values for $\chi_{1,{\rm p}}$ and $U_{\rm p}$ as a function of $L$.}
\end{table}

If our theory is asymptotically free we expect that $M_b = \Lambda \exp(-\beta/U_{\rm p})$. Further, since $\chi_{1,{\rm p}}$ is dimensionless it is expected to grow logarithmically in the continuum limit. Thus, for sufficiently large values of $L$ we expect
\begin{equation}
\chi_{1, {\rm p}} \ =\ \alpha \log(\Lambda_1 L), \ \ \ \ 
U_{\rm p} = \frac{\beta}{\log(\Lambda_2 L)},
\label{upfit}
\end{equation}
In Fig.~\ref{fig7} we show that our results are consistent with both these expectations. The parameters obtained from the fit to our data gives $\alpha =  1.77(4), \beta = 1.33(4)$, $\Lambda_1 = 0.20(1)$ and $\Lambda_2 = 0.88(9)$ \cite{Suppmat}. While in principle the value of $\beta$ can be matched to one loop perturbation theory, this can be very difficult and can require one to study extremely large correlation lengths \cite{PhysRevLett.80.1742}. Here we only study the qualitative exponential scaling of the mass gap, as was done long ago in lattice gauge theory but without using weak coupling expansion \cite{PhysRevLett.45.313}.

\section{Conclusions}

In this work we have shown how weak coupling Feynman diagrams can contain the information of a non-perturbative mass gap in an asymptotically free theory. Using a specific lattice model we first tamed the infrared divergences in the usual perturbation theory by formulating the problem in a finite volume. We then showed that the physics of the mass gap arises at sufficiently large volumes when we sample Feynman diagrams containing a finite density of interactions. The infrared divergences of the original perturbative expansion seem to be hidden in a few statistically insignificant vertex configurations. Our work suggests that a perturbative expansion organized in terms of Feynman diagrams containing a fixed density of interactions may be worth exploring. Exploring extensions of our work to gauge theories would also be interesting.

\section*{Acknowledgments}

We thank T.~Bhattacharya, S.~Hands, R.~Narayanan, U.-J.~Wiese and U.~Wolff for helpful discussions. The material presented here is based upon work supported by the U.S. Department of Energy, Office of Science, Nuclear Physics program under Award Number DE-FG02-05ER41368.

\bibliography{ref,diagmc}

\begin{thebibliography}{35}
\expandafter\ifx\csname natexlab\endcsname\relax\def\natexlab#1{#1}\fi
\expandafter\ifx\csname bibnamefont\endcsname\relax
  \def\bibnamefont#1{#1}\fi
\expandafter\ifx\csname bibfnamefont\endcsname\relax
  \def\bibfnamefont#1{#1}\fi
\expandafter\ifx\csname citenamefont\endcsname\relax
  \def\citenamefont#1{#1}\fi
\expandafter\ifx\csname url\endcsname\relax
  \def\url#1{\texttt{#1}}\fi
\expandafter\ifx\csname urlprefix\endcsname\relax\def\urlprefix{URL }\fi
\providecommand{\bibinfo}[2]{#2}
\providecommand{\eprint}[2][]{\url{#2}}

\bibitem[{\citenamefont{Kogut}(1979)}]{Kogut:1979wt}
\bibinfo{author}{\bibfnamefont{J.~B.} \bibnamefont{Kogut}},
  \bibinfo{journal}{Rev. Mod. Phys.} \textbf{\bibinfo{volume}{51}},
  \bibinfo{pages}{659} (\bibinfo{year}{1979}).

\bibitem[{\citenamefont{Prokof'ev and Svistunov}(1998)}]{Pro98}
\bibinfo{author}{\bibfnamefont{N.~V.} \bibnamefont{Prokof'ev}}
  \bibnamefont{and} \bibinfo{author}{\bibfnamefont{B.~V.}
  \bibnamefont{Svistunov}}, \bibinfo{journal}{Phys. Rev. Lett.}
  \textbf{\bibinfo{volume}{81}}, \bibinfo{pages}{2514} (\bibinfo{year}{1998}).

\bibitem[{\citenamefont{Prokof'ev and Svistunov}(2007)}]{PhysRevLett.99.250201}
\bibinfo{author}{\bibfnamefont{N.}~\bibnamefont{Prokof'ev}} \bibnamefont{and}
  \bibinfo{author}{\bibfnamefont{B.}~\bibnamefont{Svistunov}},
  \bibinfo{journal}{Phys. Rev. Lett.} \textbf{\bibinfo{volume}{99}},
  \bibinfo{pages}{250201} (\bibinfo{year}{2007}).

\bibitem[{\citenamefont{Boninsegni et~al.}(2006)\citenamefont{Boninsegni,
  Prokof'ev, and Svistunov}}]{Bon06}
\bibinfo{author}{\bibfnamefont{M.}~\bibnamefont{Boninsegni}},
  \bibinfo{author}{\bibfnamefont{N.~V.} \bibnamefont{Prokof'ev}},
  \bibnamefont{and} \bibinfo{author}{\bibfnamefont{B.~V.}
  \bibnamefont{Svistunov}}, \bibinfo{journal}{Phys. Rev. E}
  \textbf{\bibinfo{volume}{74}}, \bibinfo{pages}{036701}
  (\bibinfo{year}{2006}).

\bibitem[{\citenamefont{Prokof'ev and Svistunov}(2008)}]{PhysRevB.77.125101}
\bibinfo{author}{\bibfnamefont{N.~V.} \bibnamefont{Prokof'ev}}
  \bibnamefont{and} \bibinfo{author}{\bibfnamefont{B.~V.}
  \bibnamefont{Svistunov}}, \bibinfo{journal}{Phys. Rev. B}
  \textbf{\bibinfo{volume}{77}}, \bibinfo{pages}{125101}
  (\bibinfo{year}{2008}).

\bibitem[{\citenamefont{Kozik et~al.}(2010)\citenamefont{Kozik, Van~Houcke,
  Gull, Pollet, Prokofev, Svistunov, and Troyer}}]{Kozik:2010zz}
\bibinfo{author}{\bibfnamefont{E.}~\bibnamefont{Kozik}},
  \bibinfo{author}{\bibfnamefont{K.}~\bibnamefont{Van~Houcke}},
  \bibinfo{author}{\bibfnamefont{E.}~\bibnamefont{Gull}},
  \bibinfo{author}{\bibfnamefont{L.}~\bibnamefont{Pollet}},
  \bibinfo{author}{\bibfnamefont{N.}~\bibnamefont{Prokofev}},
  \bibinfo{author}{\bibfnamefont{B.}~\bibnamefont{Svistunov}},
  \bibnamefont{and} \bibinfo{author}{\bibfnamefont{M.}~\bibnamefont{Troyer}},
  \bibinfo{journal}{Europhys. Lett.} \textbf{\bibinfo{volume}{90}},
  \bibinfo{pages}{10004} (\bibinfo{year}{2010}).

\bibitem[{\citenamefont{Cohen et~al.}(2015)\citenamefont{Cohen, Gull, Reichman,
  and Millis}}]{PhysRevLett.115.266802}
\bibinfo{author}{\bibfnamefont{G.}~\bibnamefont{Cohen}},
  \bibinfo{author}{\bibfnamefont{E.}~\bibnamefont{Gull}},
  \bibinfo{author}{\bibfnamefont{D.~R.} \bibnamefont{Reichman}},
  \bibnamefont{and} \bibinfo{author}{\bibfnamefont{A.~J.}
  \bibnamefont{Millis}}, \bibinfo{journal}{Phys. Rev. Lett.}
  \textbf{\bibinfo{volume}{115}}, \bibinfo{pages}{266802}
  (\bibinfo{year}{2015}).

\bibitem[{\citenamefont{Buividovich}(2016)}]{Buividovich:2015qba}
\bibinfo{author}{\bibfnamefont{P.~V.} \bibnamefont{Buividovich}},
  \bibinfo{journal}{PoS} \textbf{\bibinfo{volume}{LATTICE2015}},
  \bibinfo{pages}{293} (\bibinfo{year}{2016}), \eprint{1510.06568}.

\bibitem[{\citenamefont{Buividovich and Davody}(2017)}]{Buividovich:2017sdn}
\bibinfo{author}{\bibfnamefont{P.~V.} \bibnamefont{Buividovich}}
  \bibnamefont{and} \bibinfo{author}{\bibfnamefont{A.}~\bibnamefont{Davody}}
  (\bibinfo{year}{2017}), \eprint{1705.03368}.

\bibitem[{\citenamefont{Kopper et~al.}(1995)\citenamefont{Kopper, Magnen, and
  Rivasseau}}]{Kopper:1993mj}
\bibinfo{author}{\bibfnamefont{C.}~\bibnamefont{Kopper}},
  \bibinfo{author}{\bibfnamefont{J.}~\bibnamefont{Magnen}}, \bibnamefont{and}
  \bibinfo{author}{\bibfnamefont{V.}~\bibnamefont{Rivasseau}},
  \bibinfo{journal}{Commun. Math. Phys.} \textbf{\bibinfo{volume}{169}},
  \bibinfo{pages}{121} (\bibinfo{year}{1995}).

\bibitem[{\citenamefont{Kopper}(1999)}]{Kopper:1999jj}
\bibinfo{author}{\bibfnamefont{C.}~\bibnamefont{Kopper}},
  \bibinfo{journal}{Commun. Math. Phys.} \textbf{\bibinfo{volume}{202}},
  \bibinfo{pages}{89} (\bibinfo{year}{1999}).

\bibitem[{\citenamefont{Gross and Neveu}(1974)}]{PhysRevD.10.3235}
\bibinfo{author}{\bibfnamefont{D.~J.} \bibnamefont{Gross}} \bibnamefont{and}
  \bibinfo{author}{\bibfnamefont{A.}~\bibnamefont{Neveu}},
  \bibinfo{journal}{Phys. Rev. D} \textbf{\bibinfo{volume}{10}},
  \bibinfo{pages}{3235} (\bibinfo{year}{1974}).

\bibitem[{\citenamefont{Witten}(1978)}]{Witten1978110}
\bibinfo{author}{\bibfnamefont{E.}~\bibnamefont{Witten}},
  \bibinfo{journal}{Nuclear Physics B} \textbf{\bibinfo{volume}{145}},
  \bibinfo{pages}{110 } (\bibinfo{year}{1978}), ISSN \bibinfo{issn}{0550-3213}.

\bibitem[{\citenamefont{Dunne and Unsal}(2012)}]{Dunne:2012ae}
\bibinfo{author}{\bibfnamefont{G.~V.} \bibnamefont{Dunne}} \bibnamefont{and}
  \bibinfo{author}{\bibfnamefont{M.}~\bibnamefont{Unsal}},
  \bibinfo{journal}{JHEP} \textbf{\bibinfo{volume}{11}}, \bibinfo{pages}{170}
  (\bibinfo{year}{2012}), \eprint{1210.2423}.

\bibitem[{\citenamefont{Dunne and Unsal}(2014)}]{Dunne:2013ada}
\bibinfo{author}{\bibfnamefont{G.~V.} \bibnamefont{Dunne}} \bibnamefont{and}
  \bibinfo{author}{\bibfnamefont{M.}~\bibnamefont{Unsal}},
  \bibinfo{journal}{Phys. Rev.} \textbf{\bibinfo{volume}{D89}},
  \bibinfo{pages}{041701} (\bibinfo{year}{2014}), \eprint{1306.4405}.

\bibitem[{\citenamefont{Cherman et~al.}(2014)\citenamefont{Cherman, Dorigoni,
  Dunne, and Unsal}}]{Cherman:2013yfa}
\bibinfo{author}{\bibfnamefont{A.}~\bibnamefont{Cherman}},
  \bibinfo{author}{\bibfnamefont{D.}~\bibnamefont{Dorigoni}},
  \bibinfo{author}{\bibfnamefont{G.~V.} \bibnamefont{Dunne}}, \bibnamefont{and}
  \bibinfo{author}{\bibfnamefont{M.}~\bibnamefont{Unsal}},
  \bibinfo{journal}{Phys. Rev. Lett.} \textbf{\bibinfo{volume}{112}},
  \bibinfo{pages}{021601} (\bibinfo{year}{2014}), \eprint{1308.0127}.

\bibitem[{\citenamefont{Ayyar and
  Chandrasekharan}(2016{\natexlab{a}})}]{PhysRevD.93.081701}
\bibinfo{author}{\bibfnamefont{V.}~\bibnamefont{Ayyar}} \bibnamefont{and}
  \bibinfo{author}{\bibfnamefont{S.}~\bibnamefont{Chandrasekharan}},
  \bibinfo{journal}{Phys. Rev. D} \textbf{\bibinfo{volume}{93}},
  \bibinfo{pages}{081701} (\bibinfo{year}{2016}{\natexlab{a}}).

\bibitem[{\citenamefont{Ayyar and Chandrasekharan}(2015)}]{PhysRevD.91.065035}
\bibinfo{author}{\bibfnamefont{V.}~\bibnamefont{Ayyar}} \bibnamefont{and}
  \bibinfo{author}{\bibfnamefont{S.}~\bibnamefont{Chandrasekharan}},
  \bibinfo{journal}{Phys. Rev. D} \textbf{\bibinfo{volume}{91}},
  \bibinfo{pages}{065035} (\bibinfo{year}{2015}).

\bibitem[{\citenamefont{Catterall}(2016)}]{Catterall2016}
\bibinfo{author}{\bibfnamefont{S.}~\bibnamefont{Catterall}},
  \bibinfo{journal}{Journal of High Energy Physics}
  \textbf{\bibinfo{volume}{2016}}, \bibinfo{pages}{121} (\bibinfo{year}{2016}).

\bibitem[{\citenamefont{Ayyar and
  Chandrasekharan}(2016{\natexlab{b}})}]{Ayyar2016}
\bibinfo{author}{\bibfnamefont{V.}~\bibnamefont{Ayyar}} \bibnamefont{and}
  \bibinfo{author}{\bibfnamefont{S.}~\bibnamefont{Chandrasekharan}},
  \bibinfo{journal}{Journal of High Energy Physics}
  \textbf{\bibinfo{volume}{2016}}, \bibinfo{pages}{58}
  (\bibinfo{year}{2016}{\natexlab{b}}), ISSN \bibinfo{issn}{1029-8479}.

\bibitem[{\citenamefont{Catterall and Schaich}(2017)}]{PhysRevD.96.034506}
\bibinfo{author}{\bibfnamefont{S.}~\bibnamefont{Catterall}} \bibnamefont{and}
  \bibinfo{author}{\bibfnamefont{D.}~\bibnamefont{Schaich}},
  \bibinfo{journal}{Phys. Rev. D} \textbf{\bibinfo{volume}{96}},
  \bibinfo{pages}{034506} (\bibinfo{year}{2017}).

\bibitem[{\citenamefont{You et~al.}(2017)\citenamefont{You, He, Xu, and
  Vishwanath}}]{You:2017ltx}
\bibinfo{author}{\bibfnamefont{Y.-Z.} \bibnamefont{You}},
  \bibinfo{author}{\bibfnamefont{Y.-C.} \bibnamefont{He}},
  \bibinfo{author}{\bibfnamefont{C.}~\bibnamefont{Xu}}, \bibnamefont{and}
  \bibinfo{author}{\bibfnamefont{A.}~\bibnamefont{Vishwanath}}
  (\bibinfo{year}{2017}), \eprint{1705.09313}.

\bibitem[{\citenamefont{BenTov}(2015)}]{BenTov:2014eea}
\bibinfo{author}{\bibfnamefont{Y.}~\bibnamefont{BenTov}},
  \bibinfo{journal}{JHEP} \textbf{\bibinfo{volume}{07}}, \bibinfo{pages}{034}
  (\bibinfo{year}{2015}), \eprint{1412.0154}.

\bibitem[{\citenamefont{Nagai and Jansen}(2006)}]{NAGAI2006325}
\bibinfo{author}{\bibfnamefont{K.-I.} \bibnamefont{Nagai}} \bibnamefont{and}
  \bibinfo{author}{\bibfnamefont{K.}~\bibnamefont{Jansen}},
  \bibinfo{journal}{Physics Letters B} \textbf{\bibinfo{volume}{633}},
  \bibinfo{pages}{325 } (\bibinfo{year}{2006}), ISSN \bibinfo{issn}{0370-2693}.

\bibitem[{\citenamefont{Gattringer et~al.}(2007)\citenamefont{Gattringer,
  Hermann, and Limmer}}]{PhysRevD.76.014503}
\bibinfo{author}{\bibfnamefont{C.}~\bibnamefont{Gattringer}},
  \bibinfo{author}{\bibfnamefont{V.}~\bibnamefont{Hermann}}, \bibnamefont{and}
  \bibinfo{author}{\bibfnamefont{M.}~\bibnamefont{Limmer}},
  \bibinfo{journal}{Phys. Rev. D} \textbf{\bibinfo{volume}{76}},
  \bibinfo{pages}{014503} (\bibinfo{year}{2007}).

\bibitem[{\citenamefont{Bietenholz et~al.}(1995)\citenamefont{Bietenholz,
  Focht, and Wiese}}]{Bietenholz:1994ur}
\bibinfo{author}{\bibfnamefont{W.}~\bibnamefont{Bietenholz}},
  \bibinfo{author}{\bibfnamefont{E.}~\bibnamefont{Focht}}, \bibnamefont{and}
  \bibinfo{author}{\bibfnamefont{U.~J.} \bibnamefont{Wiese}},
  \bibinfo{journal}{Nucl. Phys.} \textbf{\bibinfo{volume}{B436}},
  \bibinfo{pages}{385} (\bibinfo{year}{1995}).

\bibitem[{\citenamefont{Ichinose and Nagao}(2000)}]{Ichinose:1999rr}
\bibinfo{author}{\bibfnamefont{I.}~\bibnamefont{Ichinose}} \bibnamefont{and}
  \bibinfo{author}{\bibfnamefont{K.}~\bibnamefont{Nagao}},
  \bibinfo{journal}{Mod. Phys. Lett.} \textbf{\bibinfo{volume}{A15}},
  \bibinfo{pages}{857} (\bibinfo{year}{2000}).

\bibitem[{\citenamefont{Aoki and Higashijima}(1986)}]{Aoki:1985jj}
\bibinfo{author}{\bibfnamefont{S.}~\bibnamefont{Aoki}} \bibnamefont{and}
  \bibinfo{author}{\bibfnamefont{K.}~\bibnamefont{Higashijima}},
  \bibinfo{journal}{Prog. Theor. Phys.} \textbf{\bibinfo{volume}{76}},
  \bibinfo{pages}{521} (\bibinfo{year}{1986}).

\bibitem[{\citenamefont{Bar et~al.}(2009)\citenamefont{Bar, Rath, and
  Wolff}}]{Bar:2009yq}
\bibinfo{author}{\bibfnamefont{O.}~\bibnamefont{Bar}},
  \bibinfo{author}{\bibfnamefont{W.}~\bibnamefont{Rath}}, \bibnamefont{and}
  \bibinfo{author}{\bibfnamefont{U.}~\bibnamefont{Wolff}},
  \bibinfo{journal}{Nucl. Phys.} \textbf{\bibinfo{volume}{B822}},
  \bibinfo{pages}{408} (\bibinfo{year}{2009}), \eprint{0905.4417}.

\bibitem[{\citenamefont{Korzec and Wolff}(2006)}]{Korzec:2006hy}
\bibinfo{author}{\bibfnamefont{T.}~\bibnamefont{Korzec}} \bibnamefont{and}
  \bibinfo{author}{\bibfnamefont{U.}~\bibnamefont{Wolff}},
  \bibinfo{journal}{PoS} \textbf{\bibinfo{volume}{LAT2006}},
  \bibinfo{pages}{218} (\bibinfo{year}{2006}), \eprint{hep-lat/0609022}.

\bibitem[{\citenamefont{Sharatchandra et~al.}(1981)\citenamefont{Sharatchandra,
  Thun, and Weisz}}]{Sharatchandra:1981si}
\bibinfo{author}{\bibfnamefont{H.}~\bibnamefont{Sharatchandra}},
  \bibinfo{author}{\bibfnamefont{H.}~\bibnamefont{Thun}}, \bibnamefont{and}
  \bibinfo{author}{\bibfnamefont{P.}~\bibnamefont{Weisz}},
  \bibinfo{journal}{Nucl.Phys.} \textbf{\bibinfo{volume}{B192}},
  \bibinfo{pages}{205} (\bibinfo{year}{1981}).

\bibitem[{\citenamefont{van~den Doel and Smit}(1983)}]{vandenDoel:1983mf}
\bibinfo{author}{\bibfnamefont{C.}~\bibnamefont{van~den Doel}}
  \bibnamefont{and} \bibinfo{author}{\bibfnamefont{J.}~\bibnamefont{Smit}},
  \bibinfo{journal}{Nucl.Phys.} \textbf{\bibinfo{volume}{B228}},
  \bibinfo{pages}{122} (\bibinfo{year}{1983}).

\bibitem[{Sup()}]{Suppmat}
\bibinfo{note}{For further details about our work and analysis we refer the
  reader to the attached supplementary material.}

\bibitem[{\citenamefont{Beard et~al.}(1998)\citenamefont{Beard, Birgeneau,
  Greven, and Wiese}}]{PhysRevLett.80.1742}
\bibinfo{author}{\bibfnamefont{B.~B.} \bibnamefont{Beard}},
  \bibinfo{author}{\bibfnamefont{R.~J.} \bibnamefont{Birgeneau}},
  \bibinfo{author}{\bibfnamefont{M.}~\bibnamefont{Greven}}, \bibnamefont{and}
  \bibinfo{author}{\bibfnamefont{U.-J.} \bibnamefont{Wiese}},
  \bibinfo{journal}{Phys. Rev. Lett.} \textbf{\bibinfo{volume}{80}},
  \bibinfo{pages}{1742} (\bibinfo{year}{1998}).

\bibitem[{\citenamefont{Creutz}(1980)}]{PhysRevLett.45.313}
\bibinfo{author}{\bibfnamefont{M.}~\bibnamefont{Creutz}},
  \bibinfo{journal}{Phys. Rev. Lett.} \textbf{\bibinfo{volume}{45}},
  \bibinfo{pages}{313} (\bibinfo{year}{1980}).

\end{thebibliography}

\onecolumngrid

\newpage
\appendix*

\begin{center}
\Large{ \bf Supplementary material}
\end{center}

Here we document our Monte Carlo data and explain our analysis in greater detail.

\section*{Observables}

In this work we measure three observables. They are the average monomer density $ \rho_m$ and the two bosonic susceptibilities $ \chi_1 $ and $ \chi_2 $. Expressions for these are given below in Eqns.(\ref{obs1},\ref{obs2},\ref{obs3})
\begin{eqnarray}
\rho_m &=& \frac{U}{L^2} \sum_x \langle \psi_x^4 \psi_x^3 \ \psi_x^2 \psi_x^1 \rangle \label{obs1}\\
\chi_1&=& \frac{1}{2} \sum_x \langle \psi_0^1 \psi_0^2 \ \psi_x^1 \psi_x^2 \rangle  \label{obs2}\\
\chi_2&=& \frac{1}{2} \sum_x \langle \psi_0^1 \psi_0^2 \ \psi_x^3 \psi_x^4 \rangle \label{obs3}
\end{eqnarray}
Here averages are defined using the usual definition 
\begin{equation}
\langle {\cal O} \rangle \ = \ \frac{1}{Z} \int [d\psi] \ {\cal O}\ \mathrm{e}^{-S(\psi)}
\end{equation}
where $Z$ is the partition function.

\section*{Testing the Monte Carlo Algorithm}

In order to test our Monte Carlo results we have obtained exact analytic expressions for our observables on small lattices. We combine contributions of configurations with the same number of monomers into a single family labeled by $k$. Thus, we can write the partition function and the three observables defined above through expressions of the form
\begin{equation}
Z \ = 4^{N}\sum_{k=0}^{N} z_k (U/4)^k,\ \ Z \rho_m \ =\ 4^N\sum_{k=0}^{N} a_{2k} (U/4)^{k},\ \ 
(Z/2) \chi_1 \ =\ 4^N\sum_{k=0}^{N} b_{k} (U/4)^{k},\ \ (Z/2) \chi_2 \ =\ 
4^N\sum_{k=0}^{N-1} c_{k} (U/4)^{k}.
\label{coef}
\end{equation}
In our model only those configurations which have an equal number of monomers on odd and even sites contribute to $Z$, $\rho_m$ and $\chi_1$. Thus, $z_k=a_k=b_k=0$ unless $k$ is even. For $\chi_2$ observable we must have one extra monomer on the even or odd site. This implies $c_k=0$ unless $k$ is odd. Expressions for these coefficients on square lattices of size $L=4$ and $L=6$ are given in Table \ref{tabL}. We have tested our Monte Carlo algorithm against these exact results for a variety of couplings. In Table~\ref{table_comparison_exact} we show the comparison of the Monte Carlo results with exact calculations. Overall the agreement is excellent and this gives us confidence that our sampling procedure is correct.

\clearpage

\begin{table}[t]
\begin{center}
\begin{tabular}{|r|r|r|r|r|}
\hline
\multicolumn{5}{|c|}{$L=4$} \\
\hline
$k$ & $z_k$ & $a_k$ & $b_k$ & $c_{k-1}$ \\
\hline
16 & $1$ & $1$ & $0$ & $12$ \\
14 & $32$ & $28$ & $4$ & $456$ \\
12 & $736$ & $552$ & $160$ & $10896$ \\
10 & $13952$ & $8720$ & $4032$ & $193632$ \\
8 & $240448$ & $120224$ & $78464$ & $2789376$ \\
6 & $3571712$ & $1339392$ & $1255424$ & $29884416$ \\
4 & $48234496$ & $12058624$ & $16515072$ & $201326592$ \\
2 & $536870912$ & $67108864$ & $167772160$ & $0$ \\
0 & $4294967296$ & $0$ & $1073741824$ & $-$ \\
\hline
\hline
\multicolumn{5}{|c|}{$L=6$} \\
\hline
$k$ & $z_k$ & $a_k$ & $b_k$ & $c_{k-1}$ \\
\hline
36 & $1.0\times 10^{0}$ &
$1.0\times 10^{0}$ & 
$0$ & 
$1.2 \times 10^{1}$ \\
34 & $7.2\times 10^{1}$ & 
$6.8\times 10^{1} $  & 
$4.0\times 10^{0} $  & 
$9.08\times 10^{2} $ \\
32 & $2.844\times 10^{3} $ & 
$2.528\times 10^{3}$ & 
$3.12\times 10^{2} $  & 
$3.6956\times 10^{4}$ \\
30 & $7.9464\times 10^{4}$ & 
$6.6220\times 10^{4}$ & 
$1.3068\times 10^{4} $  & 
$1.053340\times 10^{6} $ \\
28 & $1.740870\times 10^{6}$ & 
$1.354010\times 10^{6} $ & 
$3.82992\times 10^{5} $  & 
$2.3353772\times 10^{7}$ \\
26 & $3.1613256\times 10^{7}$ & 
$2.2831796\times 10^{7}$ & 
$8.728764\times 10^{6}$  & 
$4.25917988\times 10^{8}$ \\
24 & $4.93206108\times 10^{8}$ & 
$3.28804072\times 10^{8}$ & 
$1.63718040\times 10^{8}$  & 
$6.615288804\times 10^{9}$ \\
22 & $6.779854296\times 10^{9}$ & 
$4.143244292\times 10^{9}$ & 
$2.618239172\times 10^{9}$  & 
$8.9563051092\times 10^{10}$ \\
20 & $8.3706541569\times 10^{10}$ & 
$4.6503634205\times 10^{10} $ & 
$3.6573649104\times 10^{10} $  & 
$1.07405534638\times 10^{12} $ \\
18 & $9.42203679280\times 10^{11}$ & 
$4.71101839640\times 10^{11} $ &
$4.53974787568\times 10^{11}$  & 
$1.15306298447\times 10^{13}$ \\
16 & $9.77835254141\times 10^{12}$ & 
$4.34593446285\times 10^{12}$ & 
$5.06840180904\times 10^{12} $  & 
$1.11456372088\times 10^{14} $ \\
14 & $9.42271778661\times 10^{13}$ & 
$3.66439025035\times 10^{13} $ & 
$5.12909753014\times 10^{13}$  & 
$9.70247343789\times 10^{14}$ \\
12 & $8.45677287995\times 10^{14}$ & 
$2.81892429332\times 10^{14}$ & 
$4.72023746127\times 10^{14} $  & 
$7.556401282\times 10^{15}$ \\
10 & $7.05940382813\times 10^{15}$ & 
$1.96094550781\times 10^{15} $ & 
$3.9435513503\times 10^{15} $  & 
$5.17246103652\times 10^{16}$ \\
8 & $5.45416462376\times 10^{16}$ & 
$1.21203658306\times 10^{16} $ & 
$2.96708502069\times 10^{16} $  & 
$2.99410799145\times 10^{17}$ \\
6 & $3.85164352266\times 10^{17} $ & 
$6.4194058711\times 10^{16}$ & 
$1.97439045062\times 10^{17}$  & 
$1.3404915712\times 10^{18}$ \\
4 & $2.4139636992\times 10^{18}$ & 
$2.682181888\times 10^{17} $ & 
$1.11848015872\times 10^{18}$  & 
$3.641856\times 10^{18} $ \\
2 & $1.2321792\times 10^{19}$ & 
$6.84544\times 10^{17} $ & 
$4.9360128\times 10^{18} $  & 
$0$ \\
0 & $4.096\times 10^{19} $ & 
$0$ & 
$1.3312\times 10^{19} $  & 
- \\
\hline
\end{tabular}
\end{center}
\caption{\label{tabL} Coefficients in the expansion (\ref{coef}) for $L=4$ and $L=6$ lattices.}
\end{table}

\begin{table}[h]
\begin{tabular}{|cc||r|r||r|r||r|r|}
\toprule
 L &     U &                     $ \rho_m$ &                         &                      $ \chi_1$ &                          &                     $ \chi_2 $  &                        \\ \hline
 &     &                     Exact &                      Monte-Carlo &                      Exact &                       Monte-Carlo &                    Exact &                     Monte-Carlo \\
\hline
 4 &   0.1 &    252.1997 $ \times 10^{-5} $ &    248(4) $ \times 10^{-5} $ &   50250.3033 $ \times 10^{-5} $ &   50244(4) $ \times 10^{-5} $ &   376.4032 $ \times 10^{-4} $ &   369(6) $ \times 10^{-4} $ \\
 4 &   0.3 &    243.6685 $ \times 10^{-4} $ &    244(1) $ \times 10^{-4} $ &     5226.738 $ \times 10^{-4} $ &    5227(1) $ \times 10^{-4} $ &  1161.9073 $ \times 10^{-4} $ &  1163(6) $ \times 10^{-4} $ \\
 4 &   0.5 &    783.7912 $ \times 10^{-4} $ &    787(2) $ \times 10^{-4} $ &    5621.6117 $ \times 10^{-4} $ &    5626(3) $ \times 10^{-4} $ &  2027.0147 $ \times 10^{-4} $ &  2034(6) $ \times 10^{-4} $ \\
 4 &   1.0 &      5000.0 $ \times 10^{-4} $ &   4996(6) $ \times 10^{-4} $ &    5499.6705 $ \times 10^{-4} $ &    5503(1) $ \times 10^{-4} $ &  2999.6705 $ \times 10^{-4} $ &  3006(3) $ \times 10^{-4} $ \\
 4 &   2.0 &   9216.2088 $ \times 10^{-4} $ &   9215(2) $ \times 10^{-4} $ &   14054.0293 $ \times 10^{-5} $ &   14059(6) $ \times 10^{-5} $ &   506.7537 $ \times 10^{-4} $ &   507(1) $ \times 10^{-4} $ \\
 4 &  20.0 &  99937.3631 $ \times 10^{-5} $ &  99933(2) $ \times 10^{-5} $ &  125156.2985 $ \times 10^{-8} $ &  125169(6) $ \times 10^{-8} $ &    46.9189 $ \times 10^{-6} $ &    50(2) $ \times 10^{-6} $ \\ 
 \hline 
 \hline
 6 &   0.1 &     271.345 $ \times 10^{-5} $ &    272(4) $ \times 10^{-5} $ &    6573.4421 $ \times 10^{-4} $ &    6577(2) $ \times 10^{-4} $ &    71.9009 $ \times 10^{-3} $ &    72(1) $ \times 10^{-3} $ \\
 6 &   0.3 &    276.4455 $ \times 10^{-4} $ &    276(1) $ \times 10^{-4} $ &    7210.1331 $ \times 10^{-4} $ &    7211(5) $ \times 10^{-4} $ &   235.3032 $ \times 10^{-3} $ &   235(1) $ \times 10^{-3} $ \\
 6 &   0.5 &   1022.8141 $ \times 10^{-4} $ &   1024(3) $ \times 10^{-4} $ &     8686.679 $ \times 10^{-4} $ &    8688(7) $ \times 10^{-4} $ &   459.9267 $ \times 10^{-3} $ &   461(1) $ \times 10^{-3} $ \\
 6 &   1.0 &   6685.3777 $ \times 10^{-4} $ &   6690(5) $ \times 10^{-4} $ &    6398.7824 $ \times 10^{-4} $ &    6398(4) $ \times 10^{-4} $ &  4016.2128 $ \times 10^{-4} $ &  4017(6) $ \times 10^{-4} $ \\
 6 &   2.0 &   9312.1586 $ \times 10^{-4} $ &   9315(2) $ \times 10^{-4} $ &   13637.6575 $ \times 10^{-5} $ &   13633(5) $ \times 10^{-5} $ &   492.2129 $ \times 10^{-4} $ &   490(1) $ \times 10^{-4} $ \\
 6 &  20.0 &  99937.4317 $ \times 10^{-5} $ &  99934(2) $ \times 10^{-5} $ &    125117.16 $ \times 10^{-8} $ &  125124(4) $ \times 10^{-8} $ &    46.9018 $ \times 10^{-6} $ &    49(1) $ \times 10^{-6} $ \\
\hline
\end{tabular}
\caption{Comparison of our observables $ \rho_m$, $ \chi_1 $ and $ \chi_2$, calculated using our Monte-Carlo alorithm against exact calculations on $L=4 $ and $ L=6$ lattices at various couplings.\label{table_comparison_exact}}
\end{table}

\clearpage

\begin{center}
\begin{table}
\begin{tabular}{| l | c | r | r | r ||l | c|  r | r | r |}
\hline
$U$ & $L$ & $\rho_m$ & $\chi_1$  & $\chi_2$ &
$U$ & $L$ & $\rho_m$ & $\chi_1$  & $\chi_2$ \\
\hline
 0.100 &  128 &  0.0027770(82) &   1.844(16) &    0.604(16) &
 0.100 &  256 &   0.002773(14) &    2.23(12) &     0.93(18) \\
 0.150 &  256 &   0.006477(23) &   2.509(34) &    1.373(50) &
 0.200 &   32 &   0.012034(17) &  1.5099(17) &   0.7047(21) \\
 0.200 &   64 &   0.012131(13) &  2.0513(69) &   1.1665(68) &
 0.200 &  128 &   0.012194(23) &   2.952(61) &    1.995(53) \\
 0.200 &  256 &   0.012204(33) &    4.12(24) &     3.20(21) &
 0.220 &  128 &   0.015255(29) &   3.529(72) &    2.640(67) \\
 0.220 &  256 &   0.015457(34) &    6.13(17) &     5.24(16) &
 0.225 &  256 &   0.016410(22) &    6.49(13) &     5.62(11) \\
 0.230 &  256 &   0.017393(28) &    6.83(12) &     6.02(11) &
 0.240 &  128 &   0.019086(48) &    4.46(11) &     3.63(10) \\
 0.240 &  256 &   0.019471(38) &   6.956(87) &    6.197(75) &
 0.250 &   32 &   0.020324(24) &  1.7789(27) &   1.0479(29) \\
 0.250 &   64 &   0.020712(22) &   2.739(12) &    1.964(12) &
 0.250 &  128 &   0.021318(37) &   4.837(55) &    4.056(51) \\
 0.250 &  256 &   0.021866(38) &   6.881(73) &    6.153(53) &
 0.260 &  128 &   0.023946(51) &   5.323(49) &    4.590(54) \\
 0.260 &  256 &   0.024463(26) &   6.740(50) &    6.041(37) &
 0.270 &  128 &   0.026845(43) &   5.642(39) &    4.933(35) \\
 0.270 &  256 &   0.027249(31) &   6.453(41) &    5.756(29) &
 0.275 &  256 &   0.028837(45) &   6.259(58) &    5.596(42) \\
 0.280 &  128 &   0.030126(48) &   5.762(28) &    5.098(23) &
 0.280 &  256 &   0.030406(40) &   6.111(33) &    5.459(22) \\
 0.290 &  128 &   0.033624(41) &   5.722(25) &    5.069(17) &
 0.290 &  256 &   0.033851(49) &   5.855(35) &    5.212(24) \\
 0.300 &   32 &   0.032967(41) &  2.2361(43) &   1.5843(45) &
 0.300 &   64 &   0.035580(51) &   4.097(14) &    3.447(14) \\
 0.300 &  128 &   0.037380(63) &   5.552(34) &    4.916(23) &
 0.300 &  256 &    0.03765(11) &   5.647(68) &    5.030(47) \\
 0.310 &   64 &   0.039875(54) &   4.331(12) &    3.714(12) &
 0.310 &  128 &   0.041590(67) &   5.304(27) &    4.707(19) \\
 0.320 &   64 &   0.044699(60) &   4.485(11) &    3.895(10) &
 0.320 &  128 &   0.046032(65) &   5.060(22) &    4.490(16) \\
 0.325 &  256 &    0.04851(10) &   4.999(42) &    4.421(29) &
 0.330 &   32 &   0.044573(60) &  2.6648(52) &   2.0622(53) \\
 0.330 &   64 &   0.049793(60) &  4.5276(95) &   3.9557(78) &
 0.340 &   64 &   0.055166(60) &  4.4714(84) &   3.9259(64) \\
 0.340 &  128 &   0.055868(86) &   4.665(16) &    4.122(11) &
 0.350 &   32 &   0.054811(78) &  2.9649(53) &   2.3973(55) \\
 0.350 &   64 &   0.060984(61) &  4.3834(80) &   3.8506(57) &
 0.360 &   32 &   0.060820(86) &  3.1037(52) &   2.5534(52) \\
 0.360 &   64 &   0.067013(63) &  4.2600(71) &   3.7430(52) &
 0.360 &  128 &   0.067305(89) &   4.319(12) &   3.8012(86) \\
 0.370 &   32 &   0.067713(95) &  3.2319(50) &   2.6996(48) &
 0.370 &   64 &   0.073507(65) &  4.1243(68) &   3.6155(47) \\
 0.380 &   32 &    0.07479(11) &  3.3127(47) &   2.7941(44) &
 0.380 &   64 &   0.080223(68) &  3.9799(59) &   3.4835(42) \\
 0.380 &  128 &    0.08030(10) &  3.9844(91) &   3.4919(66) &
 0.390 &   32 &    0.08268(12) &  3.3511(43) &   2.8495(38) \\
 0.390 &   64 &   0.087164(70) &  3.8349(54) &   3.3484(39) &
 0.400 &   32 &    0.09106(11) &  3.3740(40) &   2.8872(34) \\
 0.400 &   64 &   0.094678(76) &  3.6880(47) &   3.2149(34) &
 0.400 &  128 &    0.09476(11) &  3.6904(68) &   3.2160(51) \\
 0.400 &  256 &    0.09494(19) &   3.695(14) &    3.220(10) &
 0.410 &   32 &    0.09942(12) &  3.3443(37) &   2.8687(30) \\
 0.420 &   32 &    0.10850(12) &  3.2971(35) &   2.8358(28) &
 0.430 &   16 &    0.09725(21) &  2.0803(27) &   1.6008(31) \\
 0.430 &   32 &    0.11723(12) &  3.2275(34) &   2.7741(26) &
 0.440 &   32 &    0.12717(13) &  3.1361(32) &   2.6953(24) \\
 0.450 &   16 &    0.11516(25) &  2.1874(26) &   1.7248(29) &
 0.450 &   32 &    0.13633(14) &  3.0491(30) &   2.6150(23) \\
 0.450 &   64 &   0.137187(94) &  3.0855(27) &   2.6525(21) &
 0.460 &   16 &    0.12600(27) &  2.2339(25) &   1.7807(28) \\
 0.470 &   16 &    0.13694(29) &  2.2660(24) &   1.8233(26) &
 0.470 &   32 &    0.15613(14) &  2.8680(27) &   2.4468(20) \\
 0.475 &   16 &    0.14269(29) &  2.2756(23) &   1.8381(25) &
 0.480 &   16 &    0.14913(30) &  2.2877(22) &   1.8560(24) \\
 0.490 &   16 &    0.16055(32) &  2.2906(21) &   1.8657(22) &
 0.500 &   16 &    0.17293(34) &  2.2890(20) &   1.8728(20) \\
 0.500 &   32 &    0.18792(16) &  2.5957(21) &   2.1941(17) &
 0.500 &   64 &    0.18760(12) &  2.6036(16) &   2.2002(13) \\
 0.500 &  128 &    0.18750(18) &  2.6050(24) &   2.2013(19) &
 0.510 &   16 &    0.18587(36) &  2.2787(19) &   1.8702(18) \\
 0.515 &   16 &    0.19203(35) &  2.2646(19) &   1.8594(17) &
 0.520 &   16 &    0.19855(36) &  2.2497(19) &   1.8477(17) \\
 0.530 &   16 &    0.21129(37) &  2.2159(18) &   1.8204(16) &
 0.550 &   16 &    0.23685(37) &  2.1356(18) &   1.7523(15) \\
 0.550 &   64 &    0.24400(14) &  2.2091(11) &  1.83022(91) &
& & & & \\
\hline
\end{tabular}
\caption{\label{full_data} Monte Carlo results for $ \rho_m $, $\chi_1$ and $\chi_2$ for various values of $U$ and $L$. }
\end{table}
\end{center}

\vskip-0.5in

\section*{Monte Carlo Results}

\begin{figure}[!htb]
\includegraphics[width=8cm]{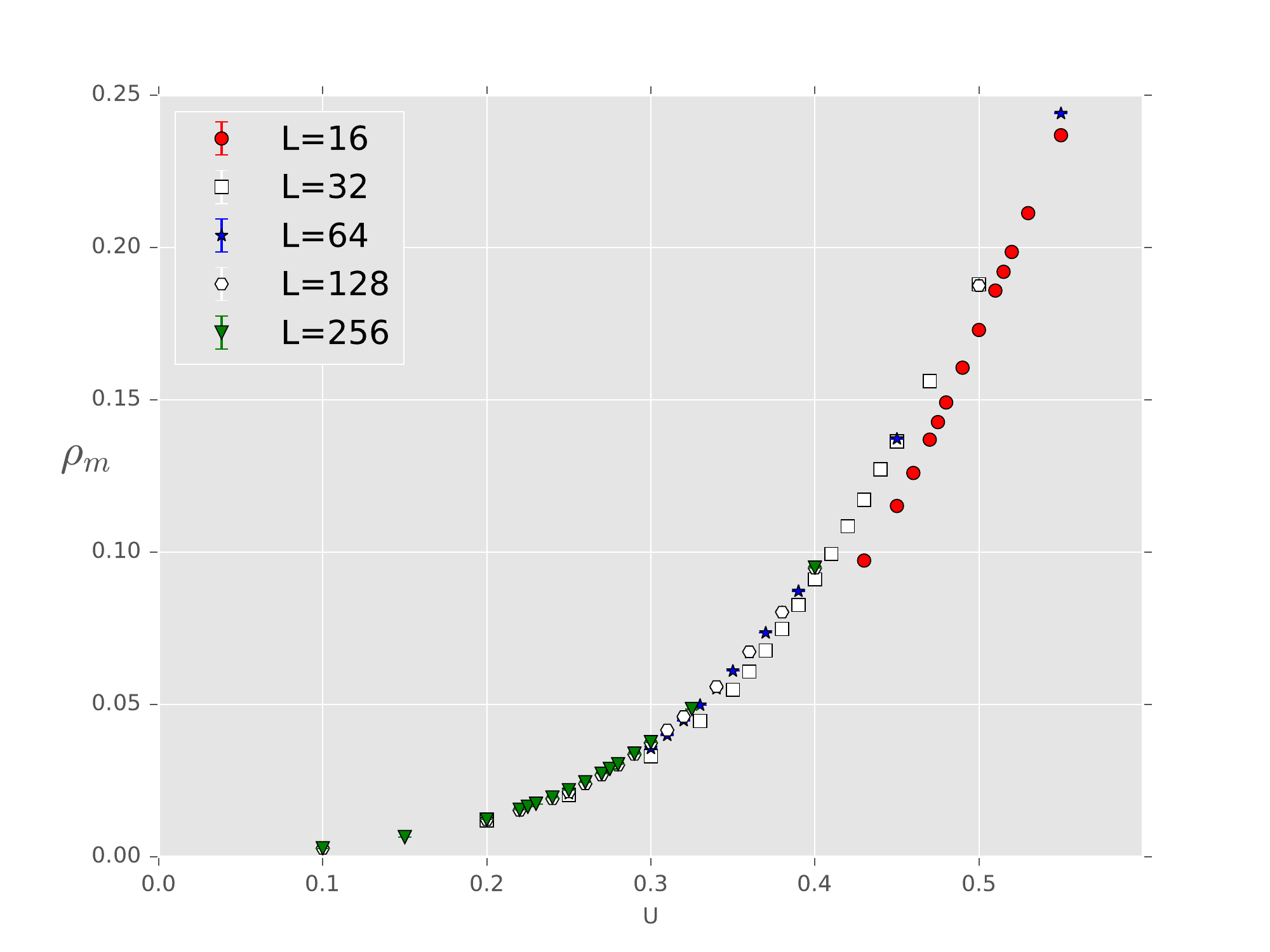}
\caption{\label{rho}Behavior of $ \rho_m $ as a function of the coupling $ U $ for the lattice sizes $L=16,32,64,128,256$.}
\end{figure}

The results from our Monte-Carlo calculations are documented in Table \ref{full_data}. In order to analyze them we first plot the behavior of $\rho_m$ as a function of the coupling $U$ for various lattice sizes in Fig. \ref{rho}. It increases from zero smoothly and even at $U=0.3$ the density is small (less than $5$ percent). Since $\rho_m$ represents the density of vertices in the diagrammatic method, it is interesting to understand its finite size scaling. We see that the curves for lattices beyond $L=32$ fall on top of each other implying that the relevant density of the vertices that play an important role in the physics is already correct at $ L=32$.

\begin{figure*}[!htb]
\includegraphics[width=7cm]{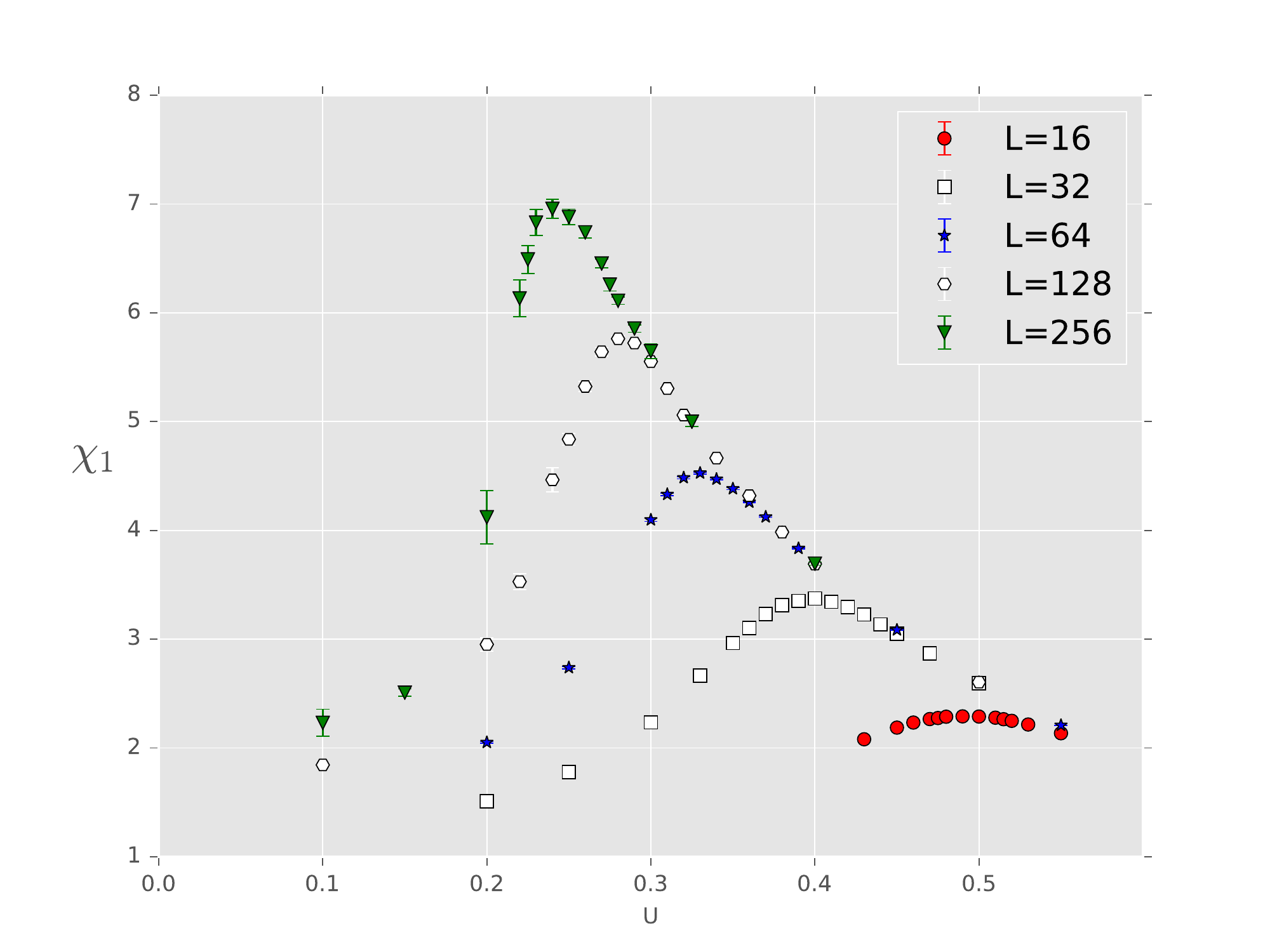}
\includegraphics[width=7cm]{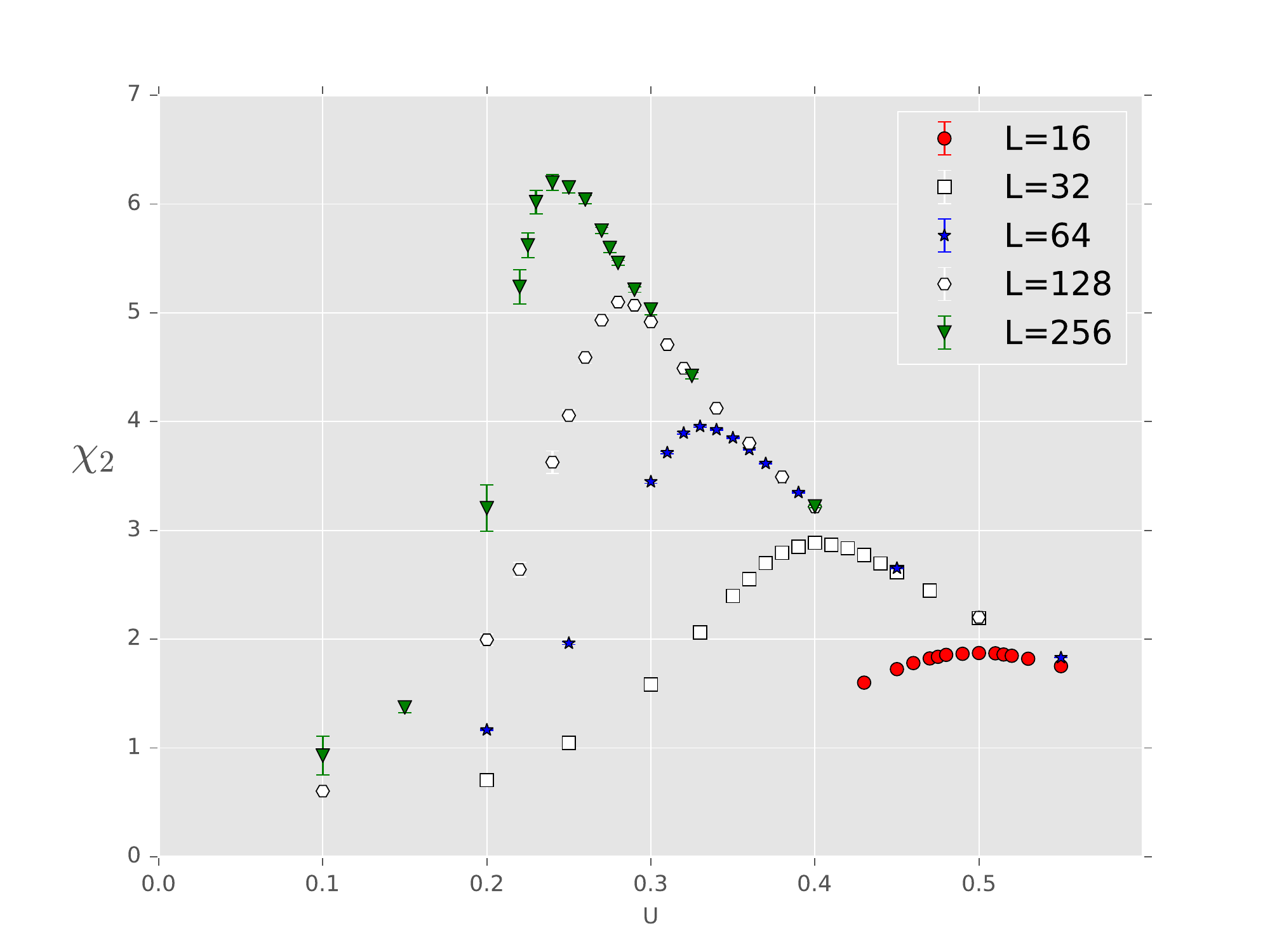}
\caption{\label{susU}Plot of the variation of the susceptibilities $ \chi_1$ $ \chi_2$ with coupling $ U $ for lattice sizes L=16,32,64,128,256. The location of the peak moves towars smaller values of $U$ as $L$ increases.} 
\end{figure*}

\begin{figure*}[!htb]
\includegraphics[width=7cm]{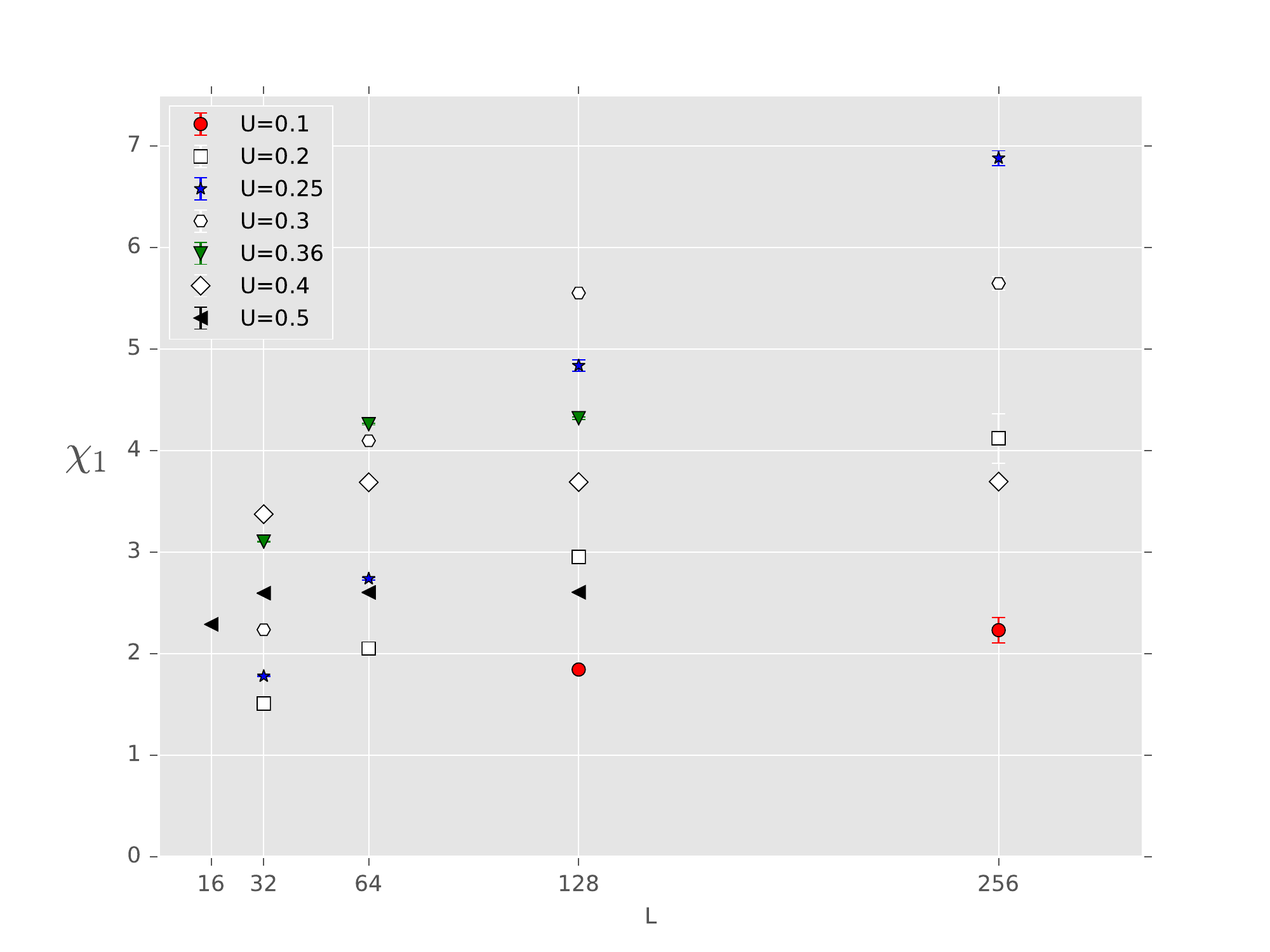}
\includegraphics[width=7cm]{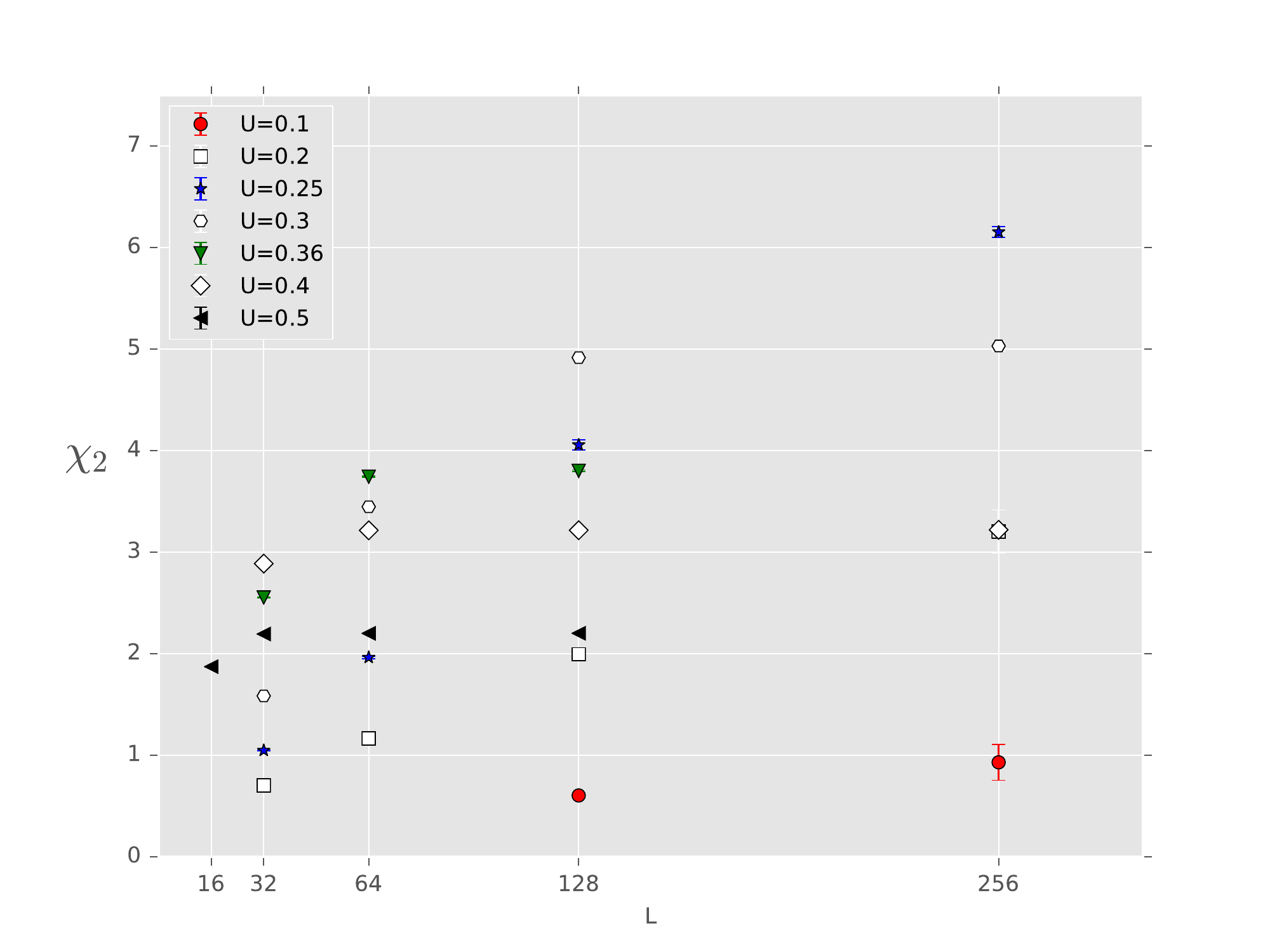}
\caption{\label{susL}Plot of the variation of the susceptibilities $ \chi_1$ $ \chi_2$ with lattice size $ L $ for various values of the coupling $ U$. The susceptibilities rise initially but finally saturate at large $L$ due to the formation of a mass gap.} 
\end{figure*}

The behavior of the susceptibilities $ \chi_1 $ and $ \chi_2 $ as a function of the coupling $ U $ for various lattice sizes is shown in Fig. \ref{susU}. In the main paper we focused on the susceptibility $\chi_1$. Here see that $\chi_2$ also shows similar behavior. We note that both susceptibilities are roughly equal and show a peak as a function of $U$ for each value of $L$. The location of the peaks can be used to define a non-perturbative mass gap. In Fig. \ref{susL} we plot these susceptibilities as a function of lattice size $L$ for various couplings $U$.

\clearpage

\clearpage

\section*{Finite size scaling to locate the critical point}

As we have seen above, both susceptibilities $\chi_1$ and $\chi_2$ show peaks for intermediate values of $U$ at a fixed value of $L$. These peaks serve as pseudo-critical points that can be defined at a lattice size $L$.  By performing a finite size scaling of how they change with lattice size one can extract the critical point of this theory. In our case we do know that the theory is asymptotically free and that the critical point is at the origin. This means we expect that 
\begin{equation}
U_p \ =\ \frac{\beta}{\log(\Lambda L)}.
\end{equation}
Our goal is to verify this qualitatively. Thus, the first step is to estimate the value of $U_p$ where the susceptibilities show a peak. Since the susceptibility is a maximum at this value of the coupling, we fit the susceptibilities near the peak to the form given by

\begin{eqnarray}
\chi_{1,2} = a_0 + a_2 \ ( U - U_p ) ^2 + a_3 \ (U-Up)^3 + a_4 \ (U-Up)^4 \label{peak_fit}
\end{eqnarray}

\begin{figure}[!htb]
\centering
\hfill
\subfloat[L=16]{\includegraphics[width=5cm]{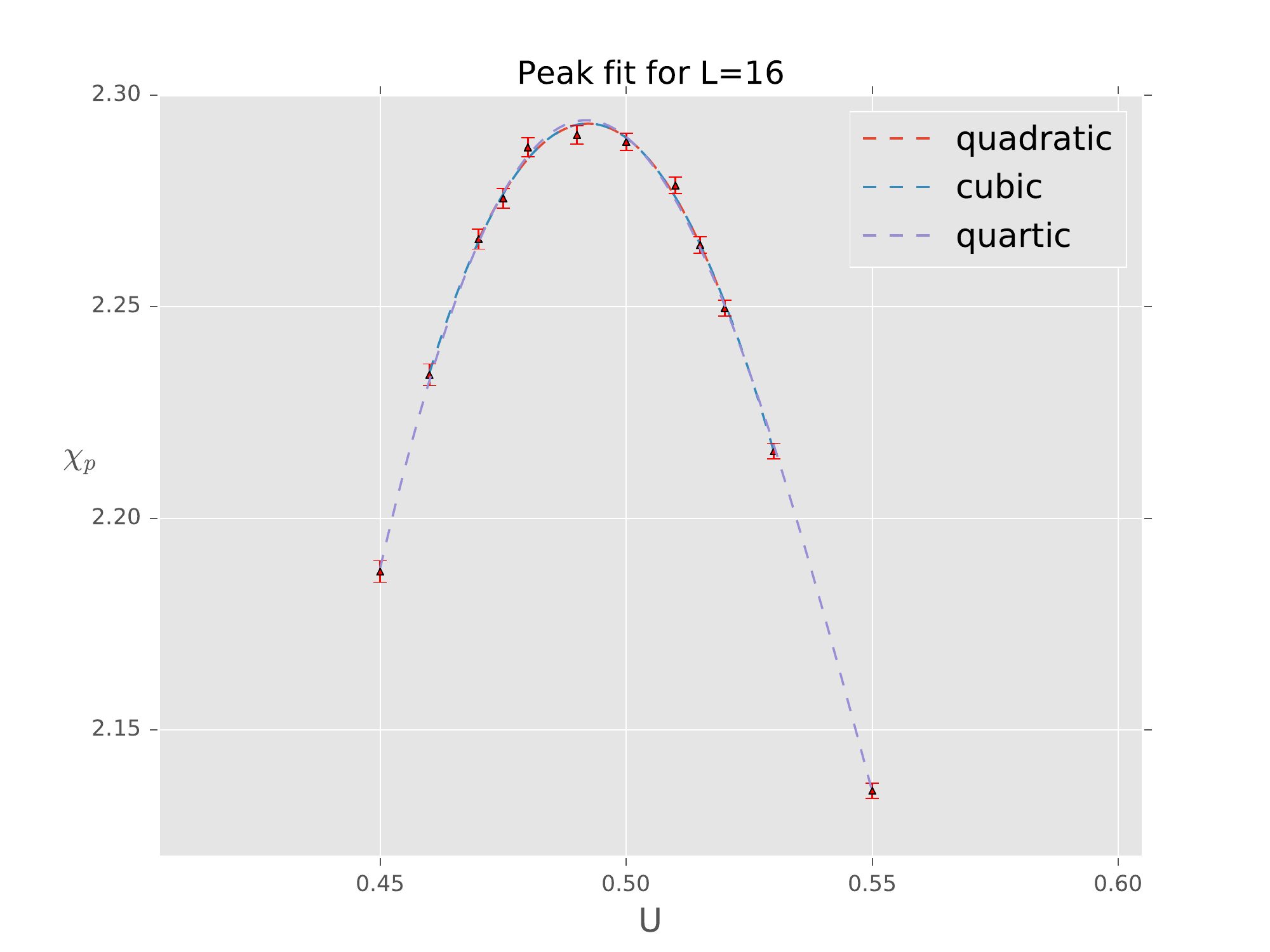}}
\subfloat[L=32]{\includegraphics[width=5cm]{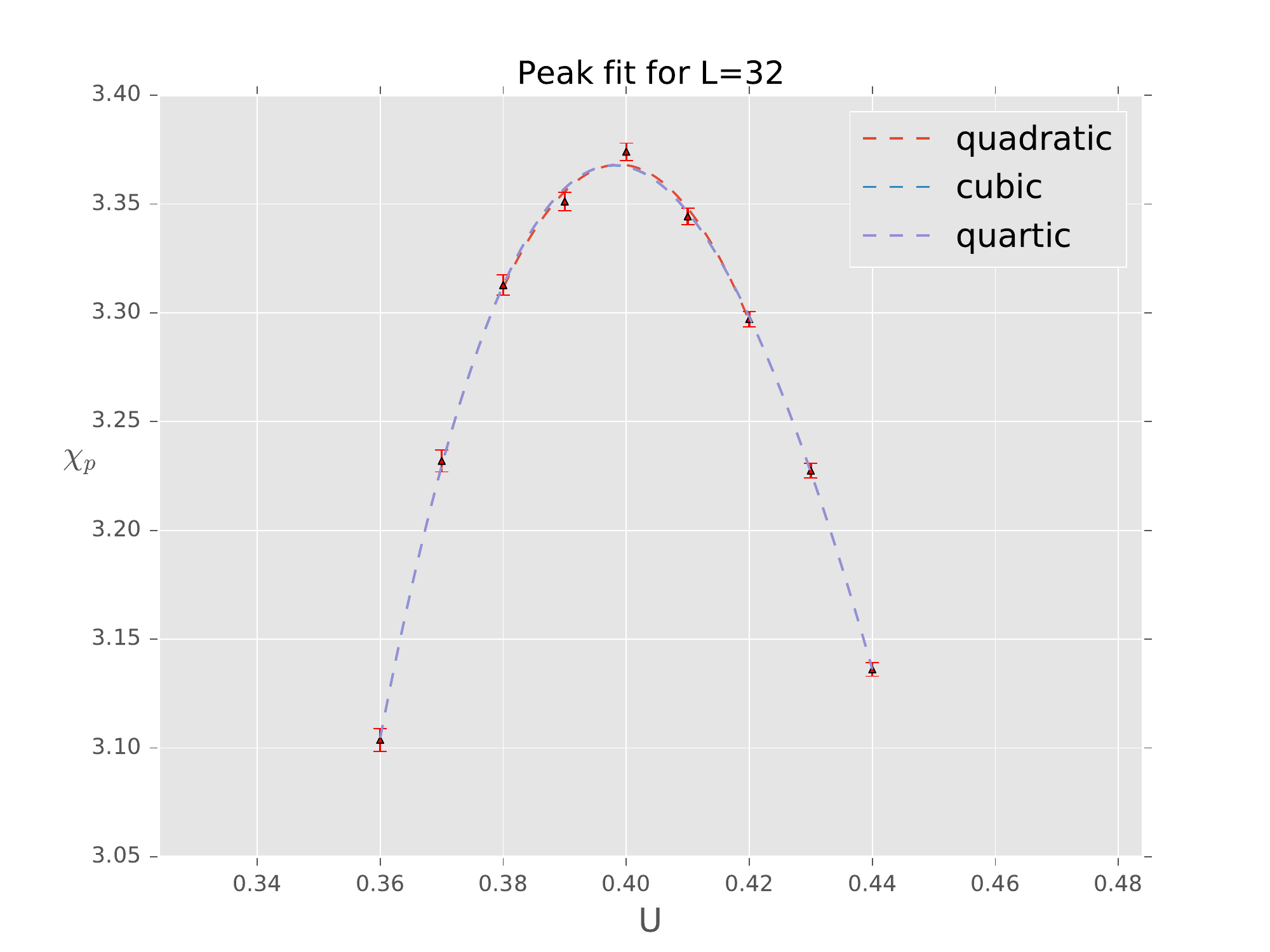}}
\subfloat[L=64]{\includegraphics[width=5cm]{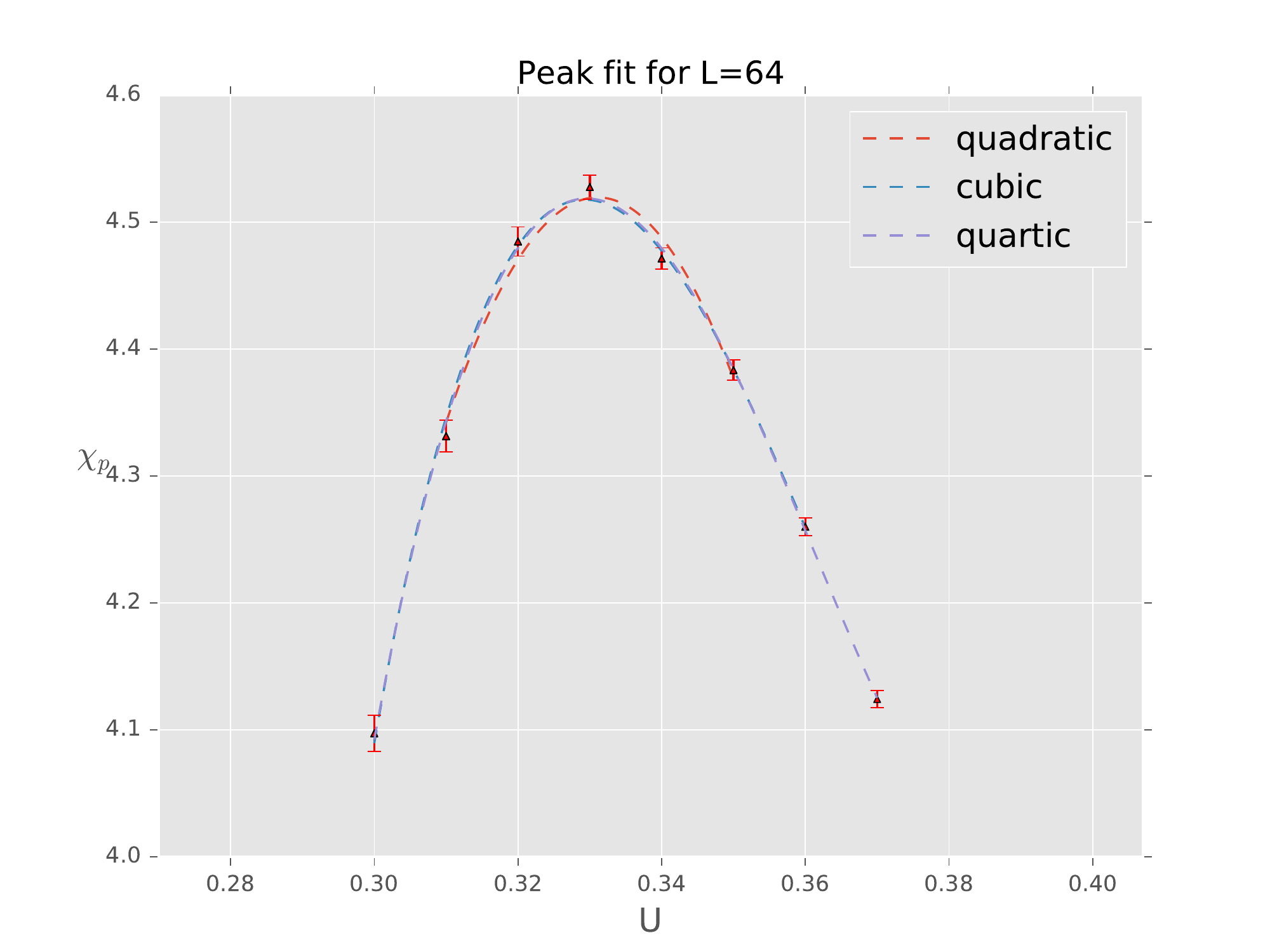}}
\newline
\subfloat[L=128]{\includegraphics[width=5cm]{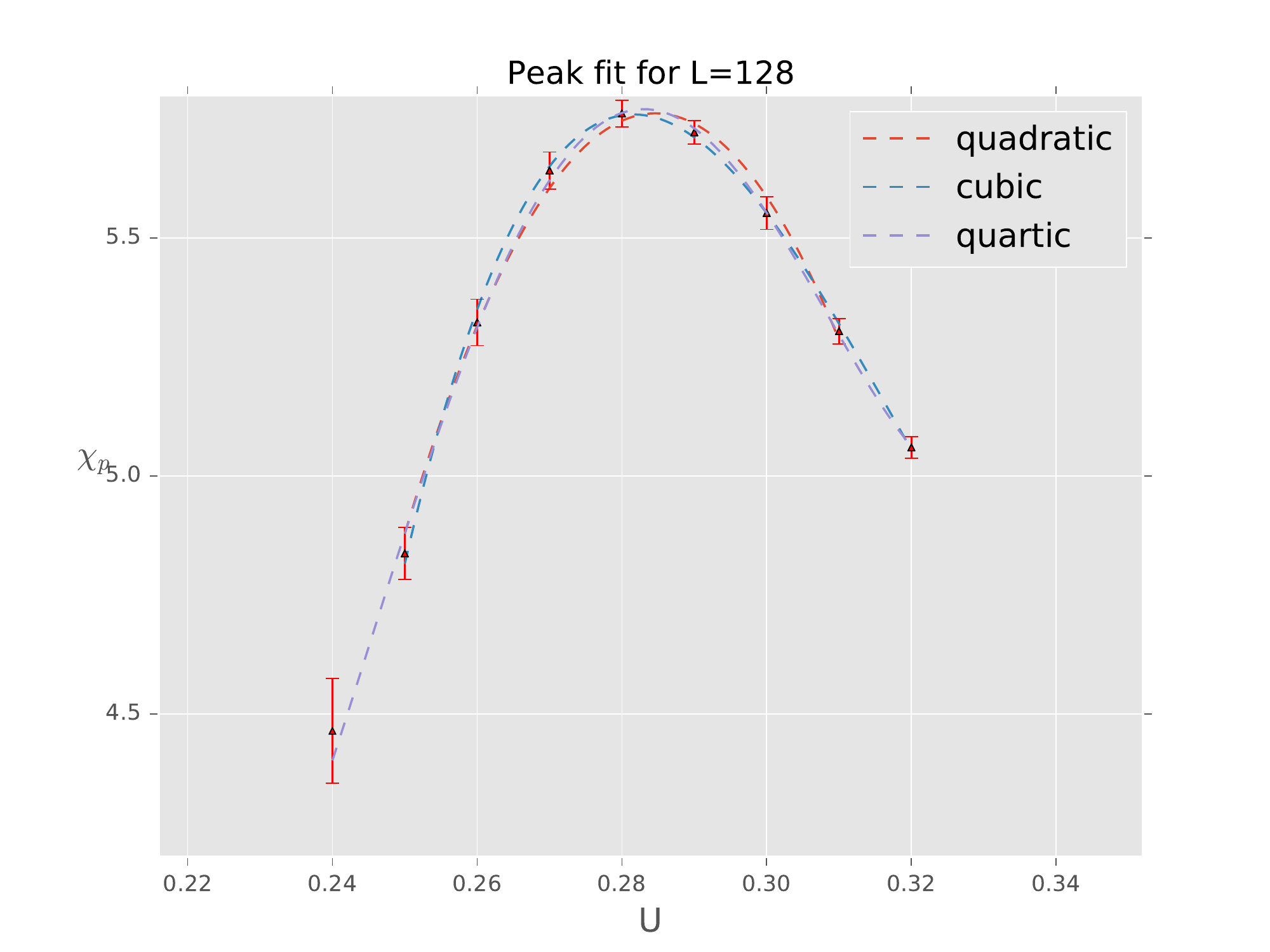}}
\subfloat[L=256]{\includegraphics[width=5cm]{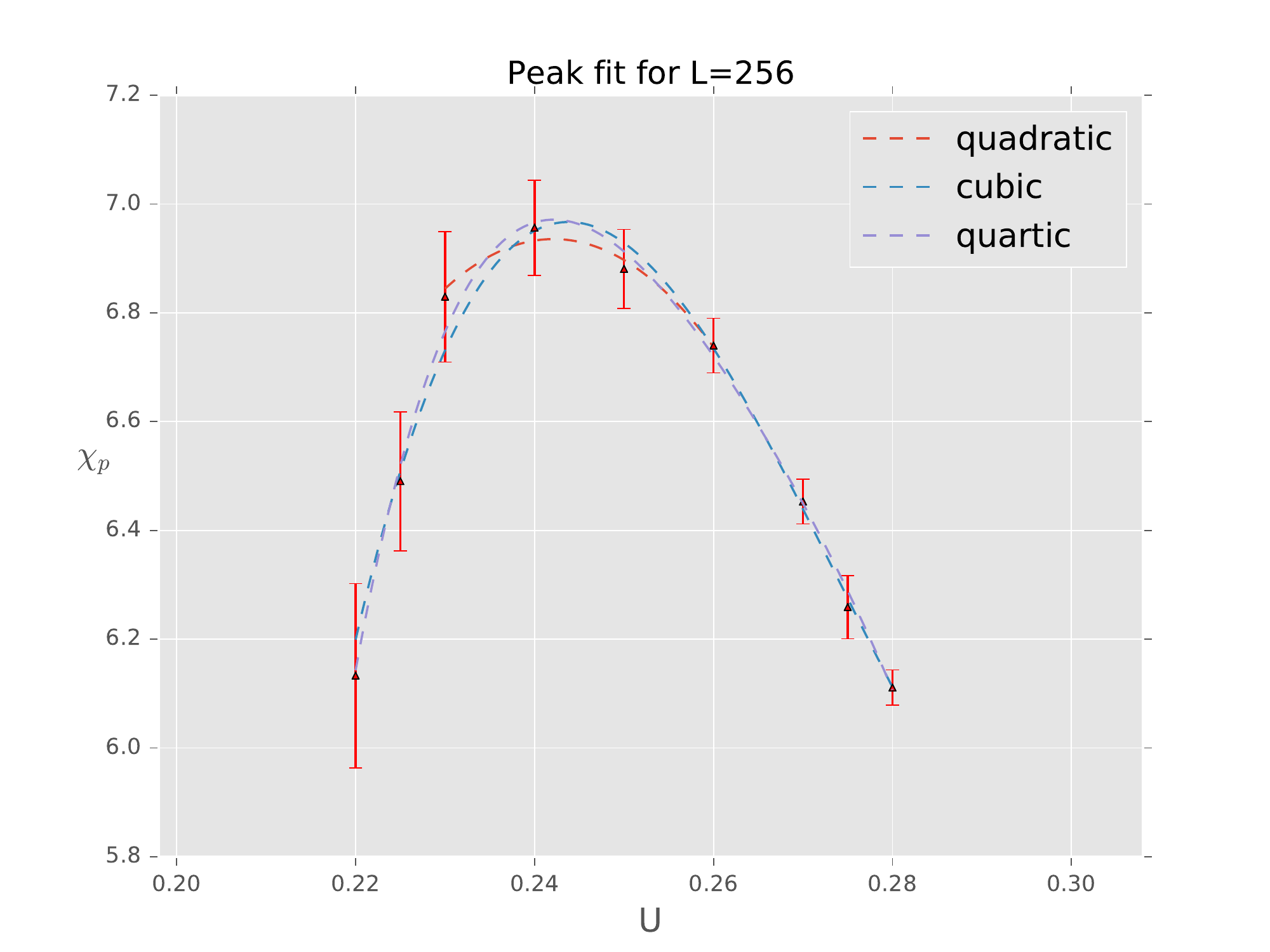}}
\caption{\label{sys_fit1_fig}Systematic fits for peak fits of $ \chi_1$ for L=16,32,64,128,256 respectively.}
\end{figure}

\begin{figure}[!htb]
\centering
\hfill
\subfloat[L=16]{\includegraphics[width=5cm]{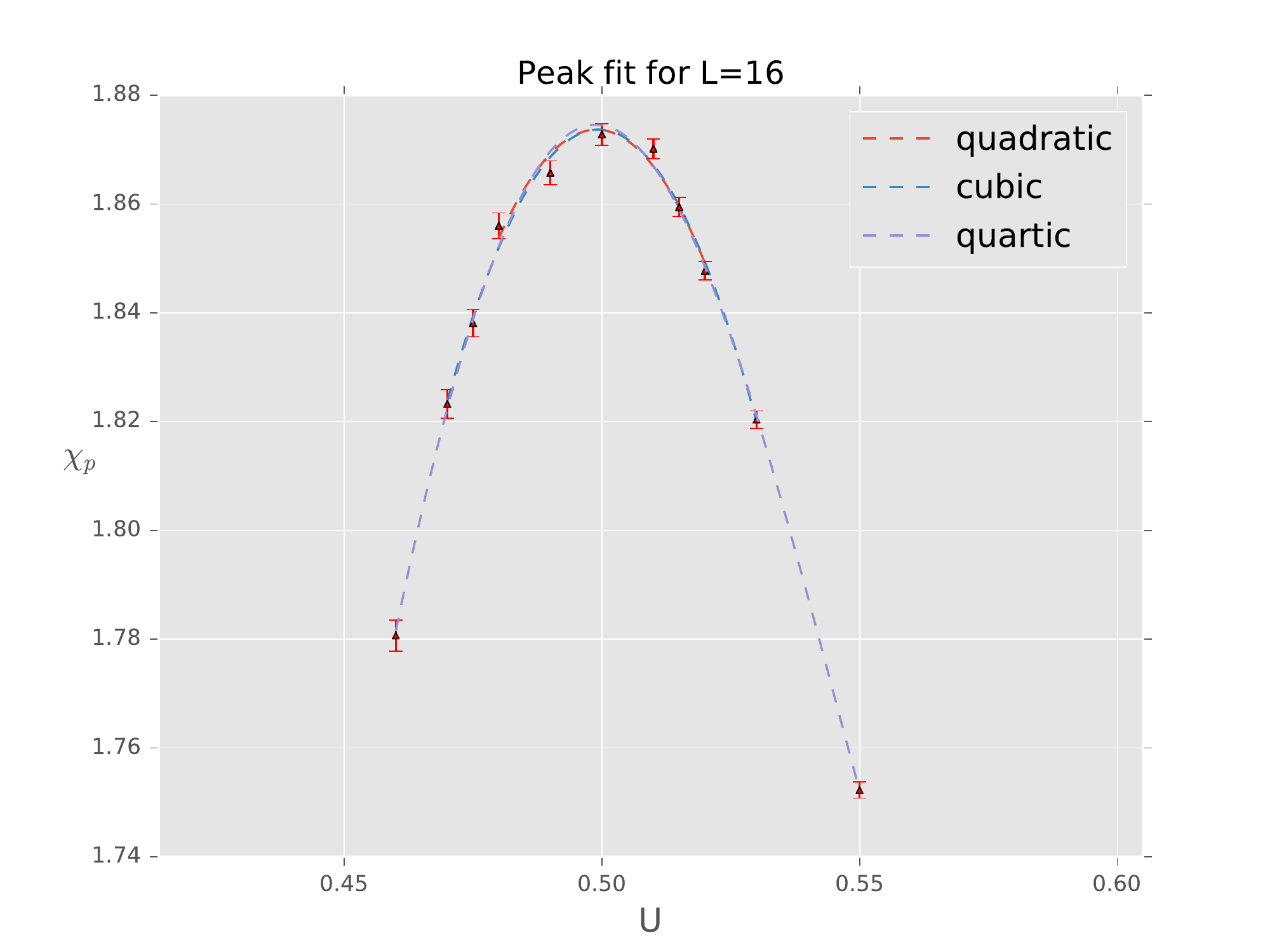}}
\subfloat[L=32]{\includegraphics[width=5cm]{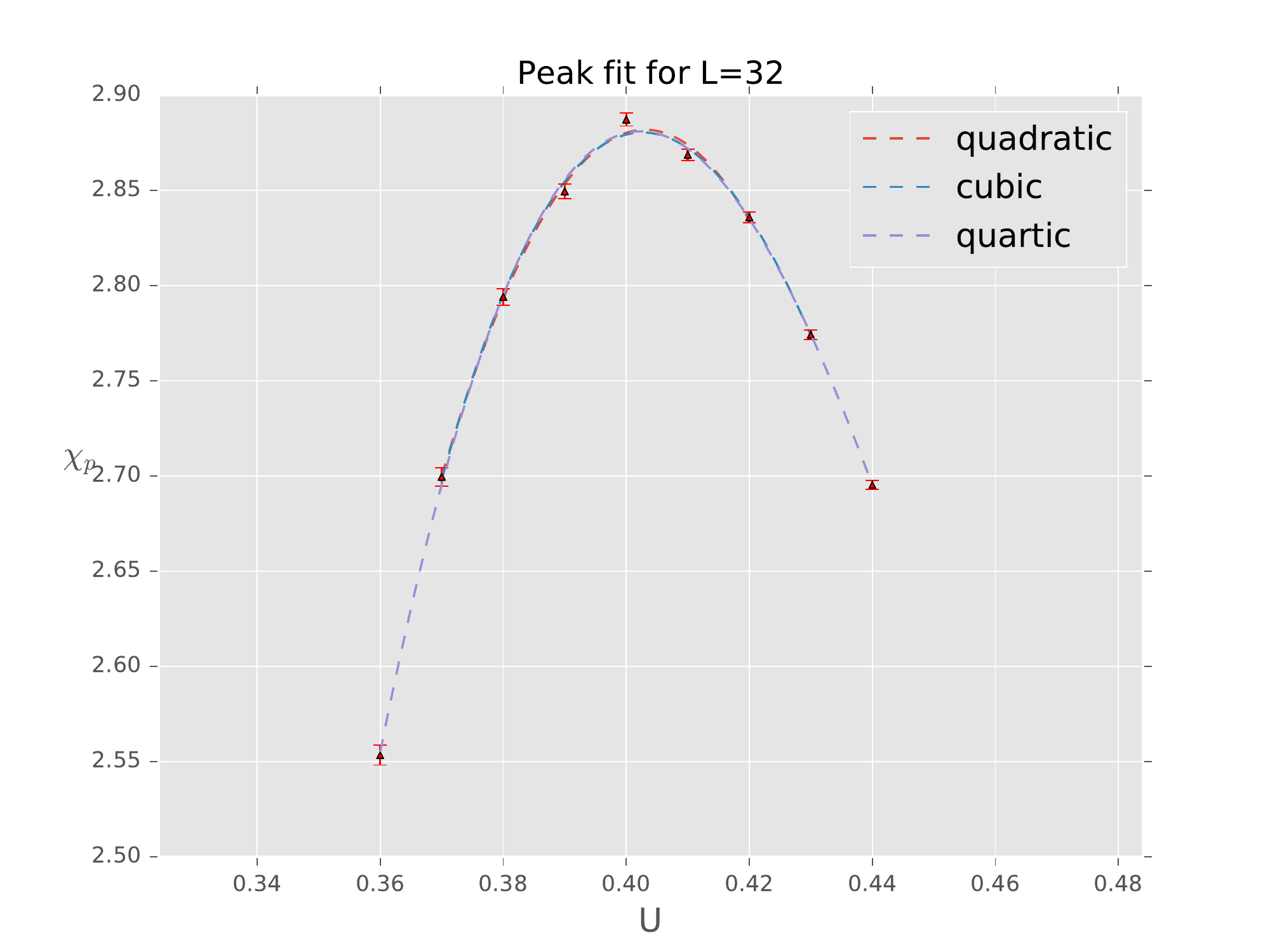}}
\subfloat[L=64]{\includegraphics[width=5cm]{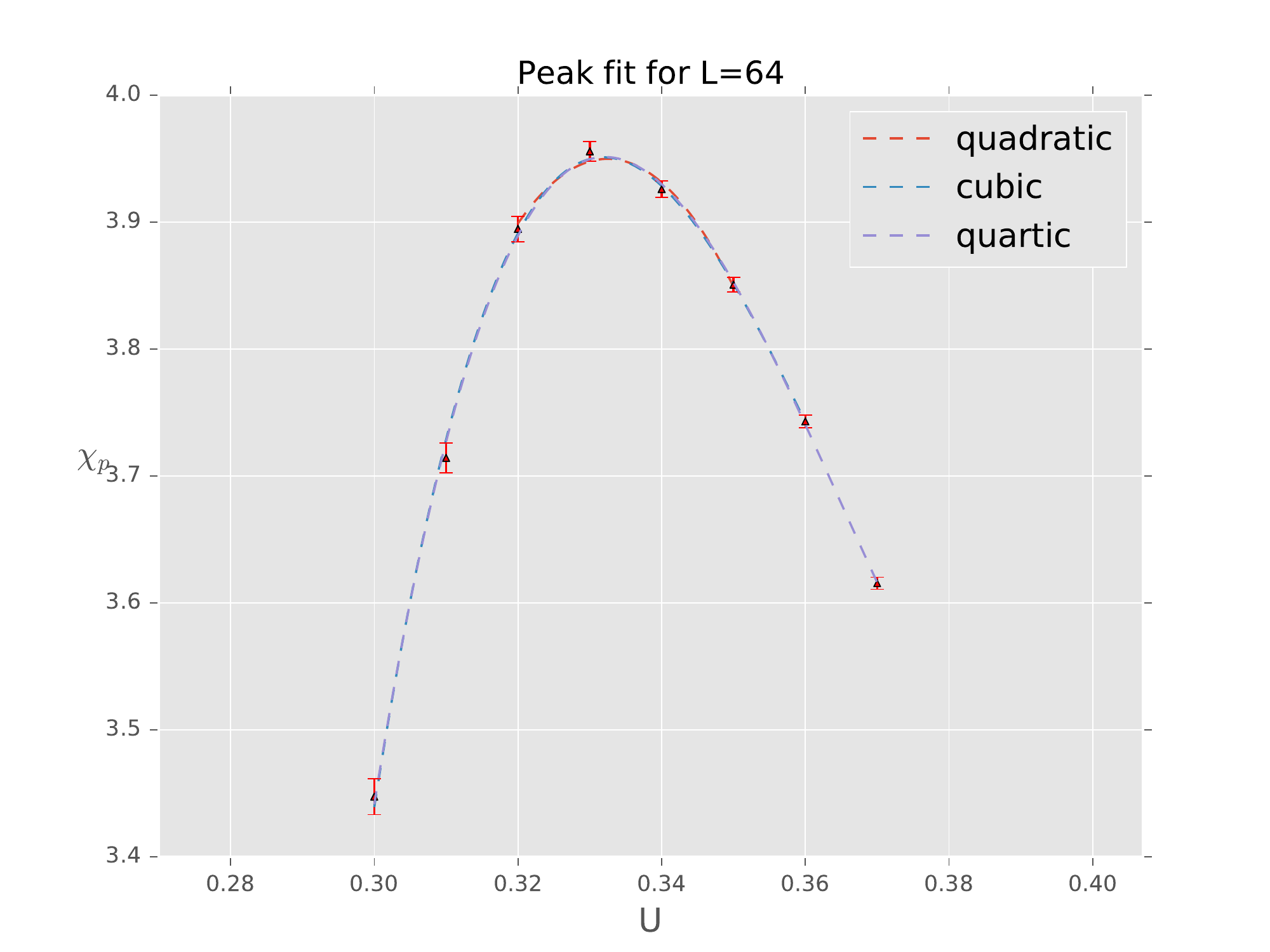}}
\newline
\subfloat[L=128]{\includegraphics[width=5cm]{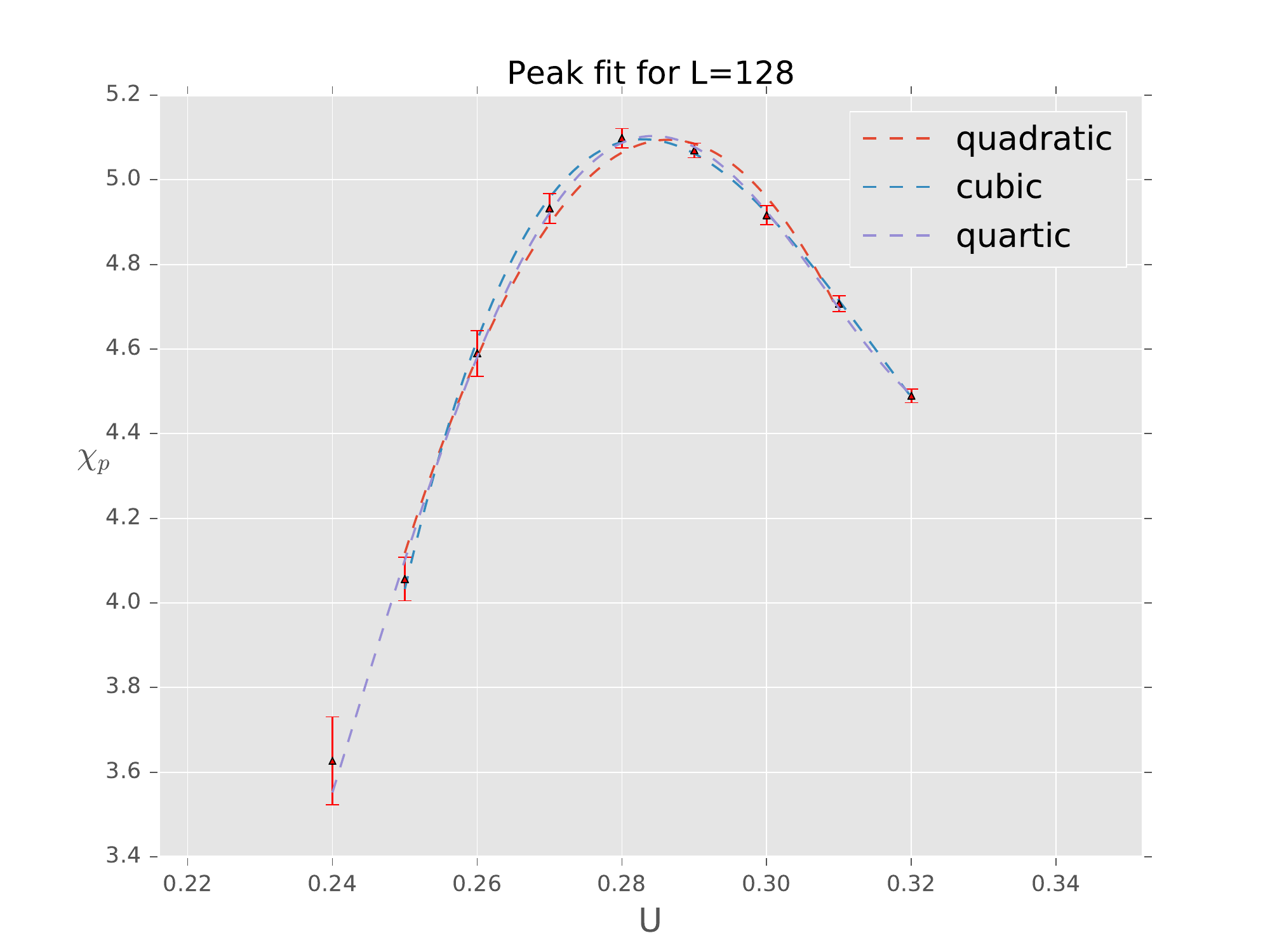}}
\subfloat[L=256]{\includegraphics[width=5cm]{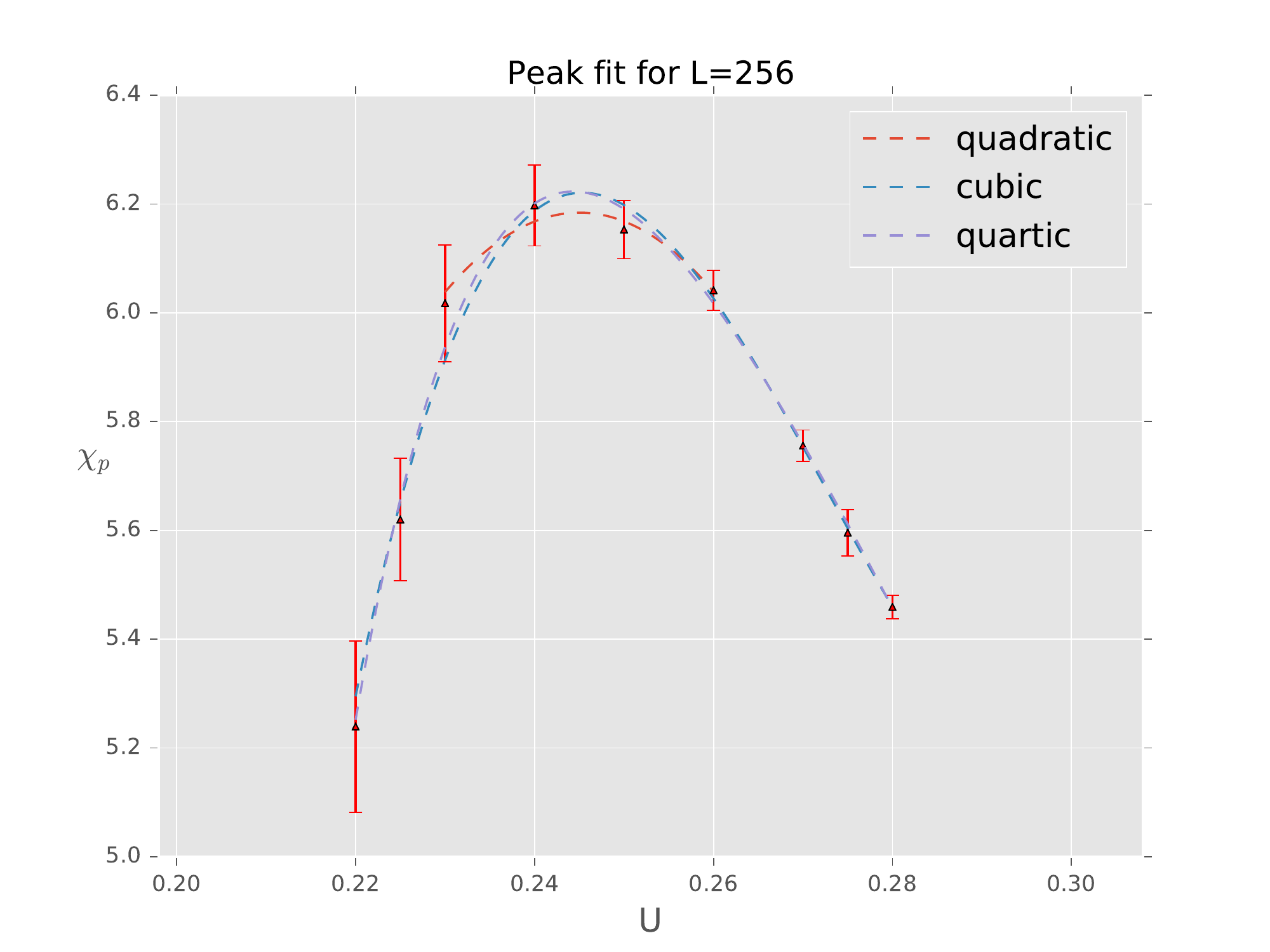}}
\caption{\label{sys_fit2_fig}Systematic fits for peak fits of $ \chi_2$ for L=16,32,64,128,256 respectively.}
\end{figure}

\begin{figure}[!htb]
\centering
\hfill
\subfloat[$\chi_1$]{\includegraphics[width=9cm]{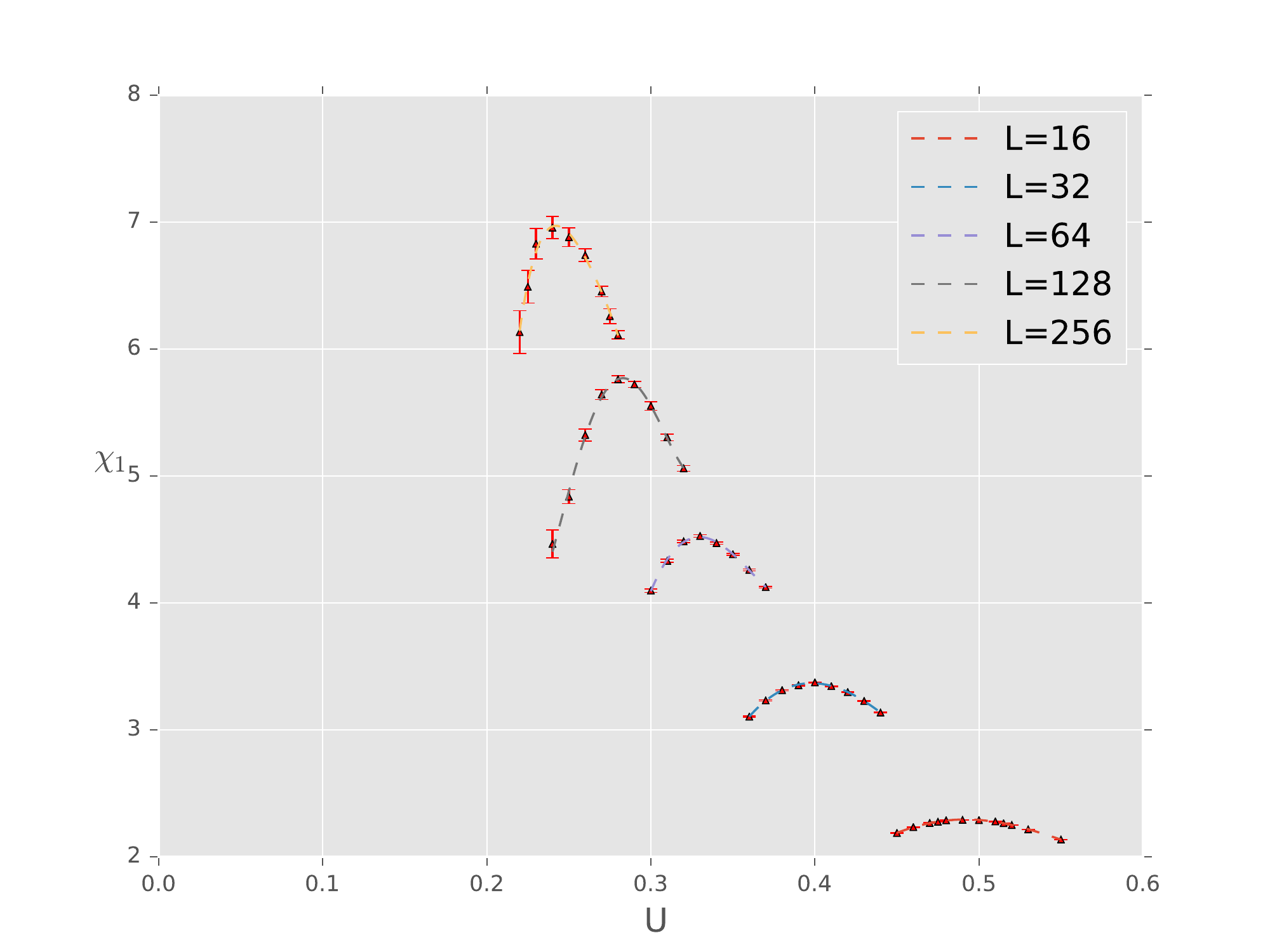}}
\subfloat[$\chi_2$]{\includegraphics[width=9cm]{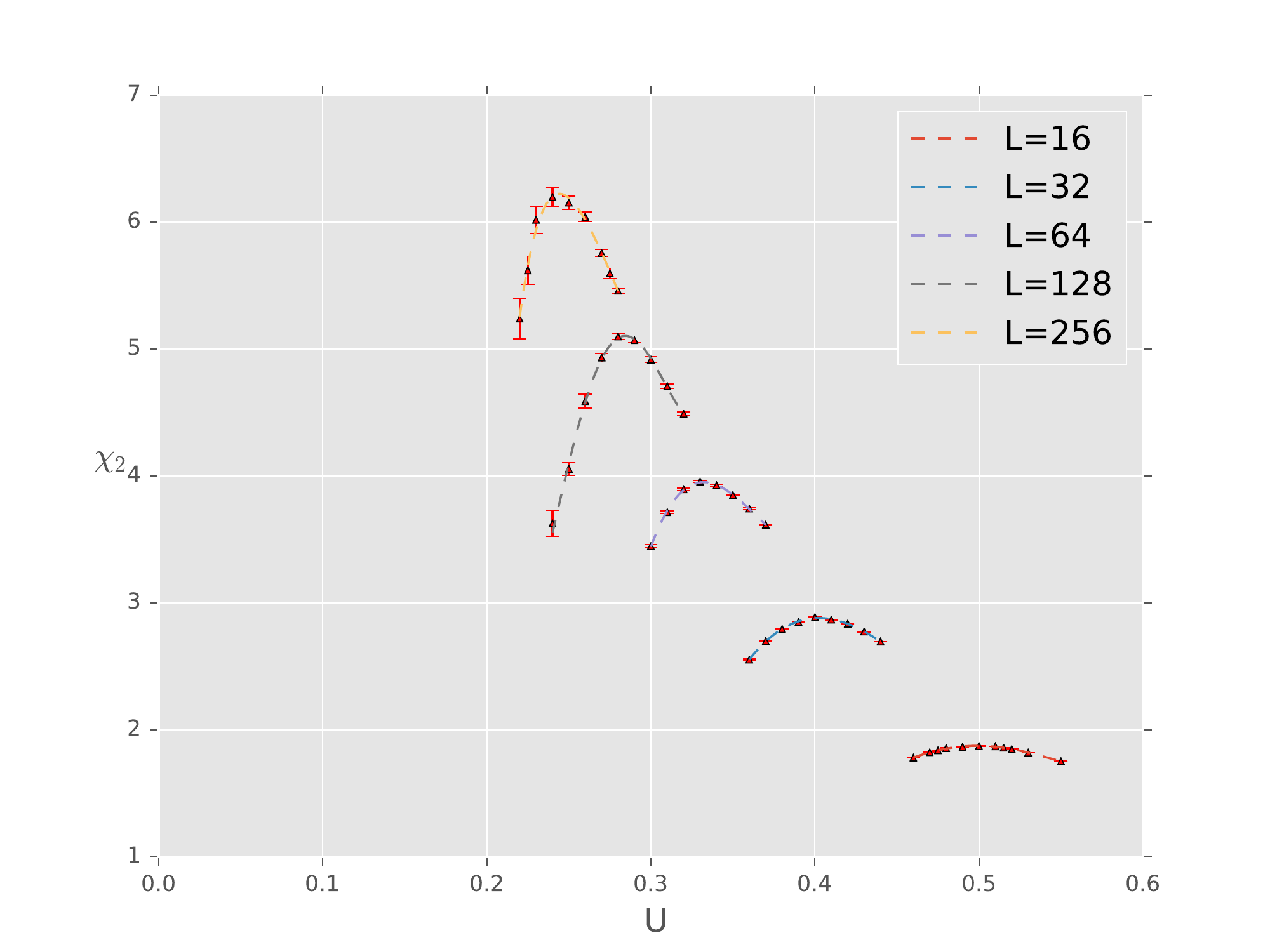}}
\caption{\label{quartic_plots}Plot for quartic fits of $ \chi_1$ $ \chi_2$  for lattice sizes L=16,32,64,128,256. It is clear that the peak of curve shifts towards smaller U as lattice size increases.}
\end{figure}

We performed a systematic analysis using three types of fits: a quadratic where $ a_3 $ and $ a_4 $ are fixed to zero, a cubic  fit where $ a_4 $ is fixed to zero and a quartic fit where all parameters are allowed to vary. Tables \ref{sys_fit1} and \ref{sys_fit2} show the results of systematic fits of $ \chi_1 $ and $ \chi_2 $ to the form given in Eqn. (\ref{peak_fit}). Figures (\ref{sys_fit1_fig}) and (\ref{sys_fit2_fig}) show these fits pictorially.  In Table \ref{best_fit} we combine all these results and quote our best estimate for the fit values of $ \chi_p $ and $ U_p $. For the errors we sum of statistical and systematic errors from the various fits. Fig. (\ref{quartic_plots}) shows the quartic fits for all the lattice sizes.

\begin{table}[hbt]
\begin{tabular}{|c|r|r|rr|rr|rr|rr|rr|}
\toprule
   L & Type of fit &  Chi-sqr &   $\chi_p$ &  $\delta \chi_p$ &      $U_p$ &    $\delta U_p$ &        $a_2$ &   $\delta a_2$ &      $a_3$ &        $\delta a_3$ &        $a_4$ &         $\delta a_4$ \\
\hline
  16 &   quadratic &   1.1114 &  2.2933 &  0.0013 &  0.4923 &  0.0005 &   -55.2534 &    3.5354 &      0.0000 &     0.0000 &       0.0000 &       0.0000 \\
  16 &       cubic &   0.9770 &  2.2933 &  0.0011 &  0.4922 &  0.0006 &   -55.7011 &    1.8922 &     36.7719 &    75.6530 &       0.0000 &       0.0000 \\
  16 &     quartic &   1.1835 &  2.2941 &  0.0012 &  0.4918 &  0.0005 &   -59.4665 &    2.7542 &    106.6690 &    45.6943 &    1897.4942 &    1222.0982 \\
\hline
  32 &   quadratic &   2.3368 &  3.3681 &  0.0042 &  0.3988 &  0.0007 &  -160.1120 &   16.5230 &      0.0000 &     0.0000 &       0.0000 &       0.0000 \\
  32 &       cubic &   1.1214 &  3.3677 &  0.0022 &  0.3980 &  0.0005 &  -158.0021 &    3.1383 &    632.6254 &   114.2130 &       0.0000 &       0.0000 \\
  32 &     quartic &   1.3987 &  3.3679 &  0.0030 &  0.3980 &  0.0006 &  -158.8703 &   10.0483 &    625.0488 &   150.5280 &     580.5512 &    6295.2304 \\
\hline
  64 &   quadratic &   3.8083 &  4.5195 &  0.0129 &  0.3311 &  0.0008 &  -398.7426 &   51.2187 &      0.0000 &     0.0000 &       0.0000 &       0.0000 \\
  64 &       cubic &   1.1521 &  4.5178 &  0.0063 &  0.3293 &  0.0007 &  -389.7572 &   17.0960 &   3783.0246 &   737.6500 &       0.0000 &       0.0000 \\
  64 &     quartic &   1.0774 &  4.5187 &  0.0064 &  0.3296 &  0.0007 &  -396.4272 &   21.9121 &   3422.7753 &   753.2829 &   10703.8631 &   23037.0197 \\
\hline
 128 &   quadratic &   0.9618 &  5.7618 &  0.0167 &  0.2846 &  0.0005 &  -737.7290 &   37.1401 &      0.0000 &     0.0000 &       0.0000 &       0.0000 \\
 128 &       cubic &   0.2793 &  5.7596 &  0.0093 &  0.2816 &  0.0005 &  -734.7338 &   24.0967 &   6693.0380 &   754.7323 &       0.0000 &       0.0000 \\
 128 &     quartic &   0.3621 &  5.7707 &  0.0113 &  0.2830 &  0.0005 &  -860.8923 &   47.0663 &   3681.3421 &   477.4259 &  151223.4532 &   31909.8240 \\
\hline
 256 &   quadratic &   0.1418 &  6.9355 &  0.0233 &  0.2421 &  0.0017 &  -620.3922 &  155.0111 &      0.0000 &     0.0000 &       0.0000 &       0.0000 \\
 256 &       cubic &   0.2982 &  6.9670 &  0.0291 &  0.2438 &  0.0007 & -1075.9692 &  100.6952 &  11689.5922 &  2632.9767 &       0.0000 &       0.0000 \\
 256 &     quartic &   0.2462 &  6.9713 &  0.0279 &  0.2421 &  0.0013 & -1101.4830 &  116.7963 &  21691.3719 &  8452.2653 & -225021.7896 &  156882.5324 \\
\hline
\end{tabular}
\caption{ Table showing the systematic peak fits for $ \chi_1 $ \label{sys_fit1}}
\end{table}

\begin{table}[htb]
\begin{tabular}{|c|r|r|rr|rr|rr|rr|rr|}
\toprule
   L & Type of fit &  Chi-sqr &   $ \chi_p$ & $\delta \chi_p$ &      $U_p$ &     $\delta U_p$ &         $a_2$ &    $\delta a_2$ &    $a_3$ &      $\delta a_3$ &      $a_4$ &          $\delta a_4$ \\
\hline
  16 &   quadratic &   2.4457 &  1.8737 &  0.0021 &  0.4989 &  0.0009 &   -55.4142 &    8.4233 &      0.0000 &     0.0000 &       0.0000 &       0.0000 \\
  16 &       cubic &   1.6920 &  1.8737 &  0.0015 &  0.4994 &  0.0010 &   -57.5439 &    2.9933 &     11.4558 &   165.1145 &       0.0000 &       0.0000 \\
  16 &     quartic &   1.6950 &  1.8746 &  0.0015 &  0.4988 &  0.0006 &   -62.2712 &    4.1086 &    125.2841 &    67.6321 &    3504.6035 &    2179.0863 \\
\hline
  32 &   quadratic &   2.8621 &  2.8818 &  0.0035 &  0.4031 &  0.0006 &  -166.1171 &   10.1122 &      0.0000 &     0.0000 &       0.0000 &       0.0000 \\
  32 &       cubic &   2.9150 &  2.8804 &  0.0033 &  0.4026 &  0.0010 &  -155.5164 &    6.4863 &    477.8623 &   405.6103 &       0.0000 &       0.0000 \\
  32 &     quartic &   2.2587 &  2.8810 &  0.0032 &  0.4023 &  0.0006 &  -159.4701 &   11.4856 &    652.8482 &   134.1990 &    2825.6431 &    6870.9043 \\
\hline
  64 &   quadratic &   1.9536 &  3.9498 &  0.0077 &  0.3325 &  0.0010 &  -327.2331 &   51.7558 &      0.0000 &     0.0000 &       0.0000 &       0.0000 \\
  64 &       cubic &   0.9398 &  3.9511 &  0.0044 &  0.3319 &  0.0005 &  -376.0841 &   11.3091 &   3977.6380 &   553.2233 &       0.0000 &       0.0000 \\
  64 &     quartic &   1.0265 &  3.9513 &  0.0049 &  0.3321 &  0.0005 &  -376.0437 &   18.9492 &   3706.3686 &   514.5751 &    1331.3217 &   18500.5385 \\
\hline
 128 &   quadratic &   2.6373 &  5.0947 &  0.0207 &  0.2864 &  0.0007 &  -736.0865 &   49.9920 &      0.0000 &     0.0000 &       0.0000 &       0.0000 \\
 128 &       cubic &   0.4076 &  5.0957 &  0.0087 &  0.2829 &  0.0005 &  -727.5055 &   23.2264 &   7660.0063 &   724.3519 &       0.0000 &       0.0000 \\
 128 &     quartic &   0.5983 &  5.1037 &  0.0109 &  0.2843 &  0.0005 &  -843.3633 &   49.1091 &   5095.0906 &   509.9798 &  142339.5815 &   34343.3067 \\
\hline
 256 &   quadratic &   0.2795 &  6.1841 &  0.0259 &  0.2450 &  0.0014 &  -647.3497 &  174.9559 &      0.0000 &     0.0000 &       0.0000 &       0.0000 \\
 256 &       cubic &   0.4180 &  6.2207 &  0.0264 &  0.2453 &  0.0007 & -1100.8041 &   90.8099 &  13474.7921 &  2443.9417 &       0.0000 &       0.0000 \\
 256 &     quartic &   0.4420 &  6.2229 &  0.0283 &  0.2443 &  0.0014 & -1096.8621 &  104.0742 &  19024.3513 &  7711.2843 & -143899.3325 &  168588.6946 \\
\hline
\end{tabular}
\caption{ Table showing the systematic peak fits for $ \chi_2 $ \label{sys_fit2}}
\end{table}

\begin{table} 
\begin{tabular}{|c|rr|rr||c|rr|rr|}
\hline
  L &  $ \chi_{1p}$ &  $ \delta \chi_{1p}$ &      $ U_{1p}$ &    $ \delta U_{1p} $ &
  L &  $ \chi_{2p}$ &  $ \delta \chi_{2p}$ &      $ U_{2p}$ &    $ \delta U_{2p} $ \\
\hline
  16 &  2.293 &   0.002 &  0.492 &  0.001 &
  16 &  1.873 &   0.003 &  0.498 &  0.002 \\
  32 &  3.368 &   0.005 &  0.398 &  0.002 &
  32 &  2.881 &   0.005 &  0.403 &  0.002 \\
  64 &  4.520 &   0.020 &  0.330 &  0.003 &
  64 &  3.950 &   0.010 &  0.332 &  0.002 \\
 128 &  5.760 &   0.030 &  0.283 &  0.003 &
 128 &  5.100 &   0.030 &  0.284 &  0.004 \\
 256 &  6.950 &   0.060 &  0.242 &  0.004 &
 256 &  6.200 &   0.070 &  0.245 &  0.004 \\
\hline
\end{tabular}
\caption{\label{best_fit} Our estimates for $ \chi_p $ and $ U_p $ for $ \chi_1 $ and $ \chi_2 $. The errors quoted combines both statistical and systematic errors.}
\end{table}

\clearpage

\section*{Asymptotic Scaling}

\begin{figure}[!htb]
\includegraphics[width=9cm]{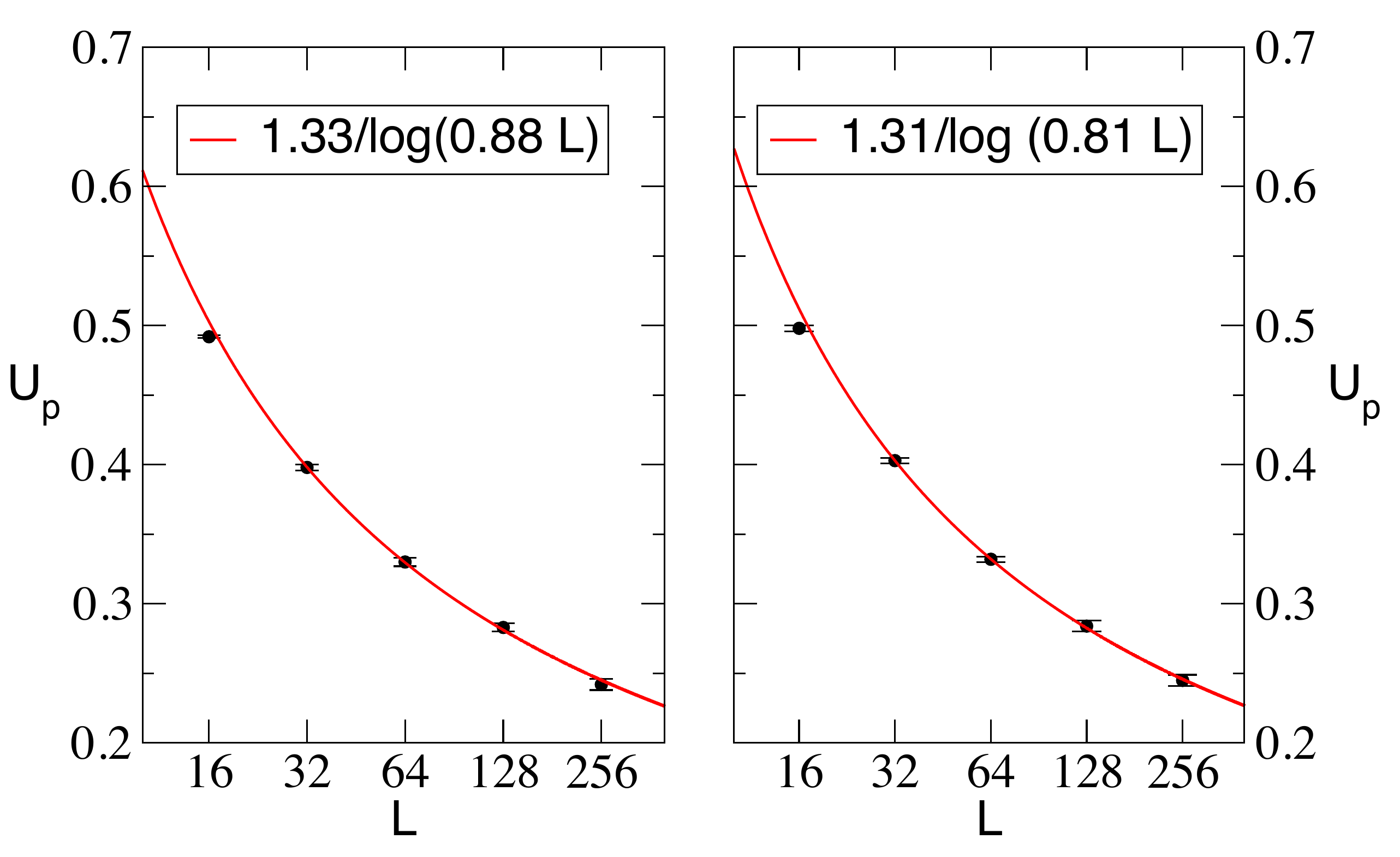}
\caption{\label{upfits} Plots of $U_p$ as a function of $L$ and the fit to the form $U_p = \beta/\log (\Lambda L)$ as discussed in the text. The left plot uses $\chi_1$ data while the right plot uses the $\chi_2$ data.}
\end{figure}

Having obtained the location of the peaks $ U_p$ as a function of $L$ for each lattice size, we proceed to check the asymptotic scaling formula through the relation
\begin{equation}
U_p= \frac{\beta}{\log(\Lambda L)},
\end{equation}
where $\lambda$ is a mass scale in lattice units. We have performed a fit of our data to this form which should be valid for sufficiently large values of $L$. If we drop the data for the lattice $ L=16$, then for $\chi_1$ we obtain $\beta = 1.33(4)$ and $\Lambda = 0.88(9)$ with a $\chi^2\mbox{/dof}=0.5$. A similar fit for $\chi_2$ gives $\beta=1.31(4)$ and $\Lambda = 0.81(8)$ with a $\chi^2/DOF = 0.1$. These fits are shown in Fig.~\ref{upfits}.

\begin{figure}[!htb]
\includegraphics[width=9cm]{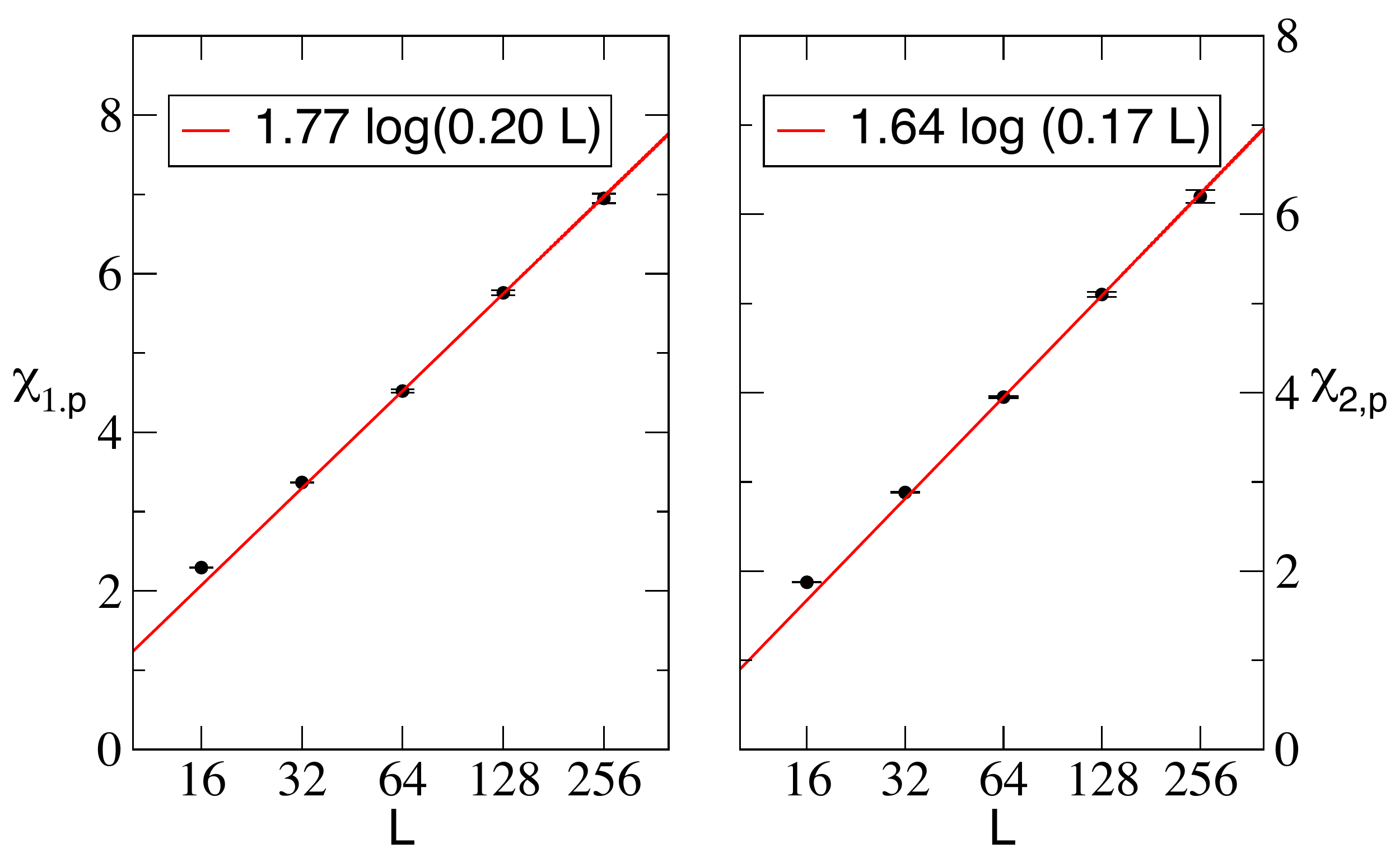}
\caption{\label{chipfits} Plots of $\chi_{1,p}$ and $\chi_{2,p}$ as a function of $L$ and the fit to the form $\chi_p = \alpha \log (\Lambda L)$ as discussed in the text.}
\end{figure}

Due to asymptotic freedom $\chi_{1,p}$ and $\chi_{2,p}$ are expected to diverge logarithmically. At leading order we expect
\begin{equation}
\chi_p \ =\ \alpha \log(\Lambda' L)
\end{equation}
where $\Lambda'$ is another mass scale in lattice units. For the fit of $\chi_{1,p}$ data we find $\alpha = 1.77(4)$ and $\Lambda' = 0.20(1)$ with a $\chi^2/DOF = 0.33$ while for the $\chi_{2,p}$ fit we find $\alpha = 1.64(4)$ and $\Lambda' = 0.17(1)$ with a $\chi^2/DOF = 0.30$. For these fits we had to drop both $L=16$ and $32$ data to obtain a good fit.  These fits data are shown in Fig.~\ref{chipfits}.

\end{document}